\documentclass[fleqn,usenatbib]{mnras}

\usepackage{newtxtext,newtxmath}

\usepackage[T1]{fontenc}

\DeclareRobustCommand{\VAN}[3]{#2}
\let\VANthebibliography\thebibliography
\def\thebibliography{\DeclareRobustCommand{\VAN}[3]{##3}\VANthebibliography}

\usepackage{graphicx}	
\usepackage{amsmath}	
\usepackage{algorithm} 
\usepackage{algpseudocode} 
\usepackage{bm}

\newcommand{\fr}[1]{{#1}}

\title[Bayesian modelling of galaxies]{Maximally informed Bayesian modelling of disc galaxies}

\author[F. Rigamonti et al.]{
Fabio Rigamonti,$^{1,2,3}$\thanks{E-mail: frigamonti@uninsubria.it}
Massimo Dotti,$^{4,2,3}$
Stefano Covino,$^{2,1}$
Francesco Haardt,$^{1,2,3}$
Marco Landoni,$^{2,1}$
\newauthor{
Walter Del Pozzo,$^{5}$
Alessandro Lupi,$^{4,3}$
Stefano Zibetti.$^{6}$}
\\
$^{1}$DiSAT, Universit\`a degli Studi dell'Insubria, via Valleggio 11, I-22100 Como, Italy\\
$^{2}$INAF, Osservatorio Astronomico di Brera, Via E. Bianchi 46, I-23807 Merate, Italy\\
$^{3}$INFN, Sezione di Milano-Bicocca, Piazza della Scienza 3, I-20126 Milano, Italy\\
$^{4}$Dipartimento di Fisica G. Occhialini, Universit\`a di Milano-Bicocca, Piazza della Scienza 3, I-20126 Milano, Italy\\
$^{5}$ Dipartimento di Fisica "E. Fermi", Università di Pisa, I-56127 Pisa, Italy and INFN, Sezione di Pisa, I-56127 Pisa, Italy\\
$^6$ INAF, Osservatorio Astrofisico di Arcetri, Largo Enrico Fermi 5, I-50125 Firenze, Italy
}

\date{Accepted 2022 May 5. Received 2022 May 2; in original form 2021 December 3}

\pubyear{2015}
\newcommand{\rowsep}[0]{\vspace{0.2cm}}
\begin{document}
\label{firstpage}
\pagerange{\pageref{firstpage}--\pageref{lastpage}}
\maketitle

\begin{abstract}
Dissecting the underlying structure of galaxies is of main importance in the framework of galaxy formation and evolution theories. While a classical bulge+disc decomposition of disc galaxies is usually taken as granted, this is only rarely solidly founded upon the full exploitation of the richness of data arising from spectroscopic studies with integral field units. In this work we describe a fully Bayesian estimation method of the global structure of disc galaxies which makes use of the wealth of photometric, kinematic, and mass-to-light ratio data, and that can be seen as a first step towards a machine-learning approach, certainly needed when dealing with larger samples of galaxies. Ours is a novel, hybrid line of action in tackling the problem of galactic parameter estimation, neither purely photometric nor orbit-based. Being rooted on a nested sampler, our code, which is available publicly as an online repository \footnotemark, allows for a statistical assessment of the need for multiple components in the dissecting process. As a first case-study the GPU-optimized code is applied to the S0 galaxy NGC-7683, finding that in this galaxy a pseudo-bulge, possibly the remnant of a bar-like structure, does exist in the center of the system. These results are then tested against the publicly available, orbit-based code DYNAMITE, finding substantial agreement.  
\end{abstract}
\footnotetext{\url{https://github.com/FabioRigamonti/BANG}}



\section{Introduction}

Since the first structural studies on galaxies, the "bulge+disc" decomposition has been a commonly adopted framework to analyse the different constituents of galaxies \citep[see e.g.][]{Mendel_2014}. Bulges were commonly thought to be spheroidal components with little net rotation well described by the de Vaucouleurs surface brightness profile \citep[][]{de-Vaucouleurs1948} originated mainly via merger events \citep[][]{Toomre_1972}. On the other hand, angular momentum conservation together with gas accretion and cooling into dark matter halos leads to the formation of a disc \citep[][]{White_1978} which has typically been described by a flat exponential profile \citep[][]{Freeman1970} supported by ordered rotation. In more recent years, though, high resolution observations have revealed that the two-component decomposition is an over-simplification of the reality.
Nowadays bulges are commonly divided into "classical bulges", which look indistinguishable from ellipticals, and "pseudobulges",  which are the result of secular processes \citep{Kormendy_2004}. The latter ones frequently show presence of disc-like features and bars resulting in an observed net rotation and/or boxy isophotes  \citep[see][for a review]{Laurikainen_2016}, this example clearly shows that a separation in bulge-plus-disc is indeed more complicated than originally thought. Attempts at modelling such a structural complexity has lead to the development of \emph{purely photometric} decomposition methods and codes, initially working on 1D azimuthally averaged surface brightness profiles \citep[e.g.,][]{Gavazzi_2000_pV}, then applied to the full 2D images \citep[e.g., GIM2D, by][]{simard_2002} and including a growing wealth of structure, such as multiple discs, bars \citep[e.g., BUDDA, by][]{budda_Gadotti}, spiral arms, rings and warps \citep[e.g.,  GALFIT, by][]{Peng_2002}.

In the last decade, the advent of integral field spectroscopy (IFS) units allowed for an unprecedentedly detailed description of the velocity field of galaxies, together with their surface brightness profile and, through the collected spectra, estimates of the mass-to-light ratio of their visible constituents. Such a richness of observational constraints is rarely fully exploited to extract information about galaxy sub-structures. 

\fr{Still, galaxy "bulge+disc" decomposition has been addressed also with IFU data providing in some cases extensions of already existing software (GALFIT) on multi-wavelength images \citep[e.g.,  BUDDI, ][]{Johnston_2017}, or combining different techniques to directly extract information on the two components from the spectra \citep[][]{Oh_2016,Tabor_2017}. These methods have been successfully applied on sub-samples of surveys such as SAMI \citep[][]{Oh_2016}, MaNGA \citep[][]{Tabor_2019} and CALIFA \citep[][]{Pak_2021}, helping to disentangle and characterise the kinematical properties of bulges and discs. In all these efforts the relative weights of the two components is determined by a purely photometric decomposition of the galaxy; this further constraint is then used to break the degeneracy, and to extract the kinematic of each component. The decomposition is generally performed in two separate steps such that the relative weights of the visible components in the potential is essentially fixed solely by the photometric decomposition  which may significantly bias the following kinematic extraction.} 

On the other hand, still triggered by IFU kinematic measurements, a different notion of structure has been put forward, which is essentially related to the dynamics of a galaxy. Starting from the seminal idea and methodology developed by \cite{Schwarzschild1979}, the stellar density field of a galaxy is seen as the superposition of suitably weighted ``orbital families'' (e.g. boxes, tubes etc.), which play the same role as bulges and discs play in the 6D phase space \citep[e.g.][]{vandeVen_2007}. In this framework, the structural decomposition of a galaxy is meant to reproduce both the measured line-of-sight velocity distribution at any projected 2D resolution element and the measured surface brightness (stellar mass surface density). The drawback of these methods is that the calculation of the orbital families is very expensive from the computational point of view, while depending on the assumed gravitational potential. Ideally one would like to fit for the gravitational potential as well as for the orbital decomposition, but this is practically impossible. Therefore the potential of the visible component is assumed at the beginning, using minimal constraints from the observed stellar surface brightness (mass density), and then is left unchanged.

In this work we develop a novel approach that conceptually sits in between the pure photometric approach and the orbit-based approach described above. As in the photometric approach, we start from the assumption that each galaxy is composed by the superposition of simple "classical" components, i.e., bulge plus one or more discs. These components are parameterized in terms of analytic density distributions and characterized by well defined dynamics in the 6D phase space, similarly to the orbital families in the orbit-based approach. Their kinematic properties, however, are not fixed by an a-priori assumed gravitational potential, but are fitted to reproduce the kinematic and photometric data simultaneously. The gravitational potential is actually an output of the fitting procedure along with the parameters that describe the different components, both from the spatial and the kinematic points of view. 

The relative light weight of the computational burden of the fitting procedure (especially relative to the orbit-based methods), allows us to adopt a fully Bayesian strategy to infer the probability distributions of the relevant model parameters.

Our goals are reached by employing the so-called nested sampling algorithm presented in \citep[][]{Skilling2004}. The architecture of our scheme allows us to explore different models until a user-defined convergence criteria is reached. Moreover we can evaluate the evidence of each dynamical prescription allowing for a direct model selection, e.g., by making it possible to check whether the inclusion of additional components/parameters is statistically significant. The computational cost of the parameter estimation is significantly reduced by parallelizing our scheme on high-performance GPU units.

The paper is organized as follows. In Section 2 we describe our methodology, in section 3 we show the application of our tool to a real galaxy and in section 4 we leave our conclusions.

\section{Methodology}
\label{sec:methodology}
In the following section we describe our methodology and its underlying assumptions regarding both the dynamical modelling and the parameter estimation. 

In our framework each galaxy is composed by the superposition of simple components (the full description of which is provided in the next sub-sections) whose parametric dependencies can be fitted to the photometric and spectroscopic data. The (semi-)analytical nature of our models reduces significantly the computational cost of the procedure, allowing for its coupling with state-of-the-art Bayesian techniques for parameter estimation and model selection. We opted for a nested sampling algorithm presented in \cite{Skilling2004}, due to its two main features: (1) it automatically evaluates the evidence\footnote{In Bayesian theory the evidence is the normalization factor of the posterior probability. The evaluation of this number is usually avoided since it requires to compute an integral over the whole parameter space.} allowing for a direct model selection, e.g. by making it possible to check whether the inclusion of additional components/parameters results (or not) in statistically preferred models; (2) it keeps on exploring the parameter space until a sufficiently large amount of models has been evaluated, stopping when the evidence accuracy reaches a user-defined threshold.

Still, the large number of model constructions and likelihood evaluations needed by the nested sampling would require more than a day on a single processor for each individual galaxy (see below for more details). 
We therefore decided to speed-up (by more than a factor of 100) the model construction by  parallelizing it on graphical processing units (GPUs). In the following we provide a complete description of all the ingredients of our algorithm, which is publicly available at \url{https://github.com/FabioRigamonti/BANG},  while the description of the parallelization strategy is discussed in Appendix~\ref{app:GPU_parallelization}.

\subsection{Extrinsic parameters: Setting the reference frame}
Each galaxy is the superposition of a spherical bulge, two exponential thin disc and a dark-matter halo.
Each model is axially symmetric and has a well defined centre, that corresponds to both the photometric and dynamical centre shared by all the components. We also assume 3 independent constant $M/L$ ratios one for the bulge and one for each of the two discs. The 2-D coordinates defining the position of the centre of the model in the plane of the sky are dubbed as $x_0$ and $y_0$, and are used to translate the observed galaxy setting its centre in the origin.

In addition to the centre, the observables associated to non-spherically symmetric components (discs) depend on two angles: the position angle P.A., i.e. the angle of the major axis of the projected discs with respect to an arbitrary direction in the plane of the sky 
\footnote{The P.A. in the model is measured starting from the horizontal direction and going counterclockwise to guarantee the alignment of the galaxy major axis with the $x$ axis after rotation.}
and the inclination angle $i$, defined as the angle between the normal to the plane of the discs\footnote{All disc components are assumed to be razor thin and to lie in the same plane.} and the line of sight (l.o.s. hereafter). The two angles allow to project the 3-D properties of the model either on the l.o.s. as well as on plane of the sky by means of two simple rotations. Hereon, we will identify with $r$ the 3-D distance of a point to the centre of the galaxy, and with $R$ the projected radius of such point on the plane of the sky. 

\fr{It is worth noting that all the assumptions we adopt in characterizing the model, such as spherical bulge, razor-like thin discs and isotropic velocity dispersion, 
are 
mandatory 
for maintaining the computational cost of the algorithm affordable. In principle, it is possible to avoid these approximations in the model construction \citep{Ciotti_2021} at the cost of a significant increase in the computational cost due to demanding numerical quadratures in solving the Jeans equations and in the line-of-sight projection \citep[see, e.g.,][]{Caravita_2021} \footnote{\fr{As an example, in the case of an exponential disc the potential itself has not a simple analytical expression outside the equatorial plane.}}. Such a computational burden is poorly compliant with Bayesian parameter estimation techniques which are necessary given the large parameter space. In the future, we plan to develop a generalised version of our model exploring different optimization strategies such as multi-GPU parallelization and machine learning techniques.}

\subsection{Intrinsic parameters: Bulge \& Halo}
\label{sec:Bulge}
We assume the bulge to be a spherically symmetric component with an isotropic velocity distribution. We implement two different potential-density pairs, an Hernquist profile (\citealt{Hernquist1990}) and a more centrally concentrated Jaffe profile \citep{Jaffe1983}. \fr{ Unlike the Hernquist profile, the Jaffe model features a density profile with a steeper inner slope $(\rho \sim r^{-2})$. Hernquist and Jaffe models can be considered representative of bulges with different inner slopes, an additional parameter that could eventually be included as a generalization of the model following, for example, \cite{Dehnen1993}.} In the following, we will describe the properties of the Hernquist model as reported in \citet{Hernquist1990}, used for the analysis discussed in the paper, while the description of the Jaffe model is reported in Appendix~\ref{app:Jaffe_profile}.
\fr{From \cite{Hernquist1990} the 3-D density profile follows}
\begin{equation}
\rho_{\rm b}(r) = \frac{M_{\rm b}}{2\upi} \frac{R_{\rm b}}{r (r + R_{\rm b})^3},
\end{equation}
where $M_{\rm b}$ is the total mass of the bulge and R$_{\rm b}$ is its scale radius.

The circular velocity of a test particle subject only to the gravitational attraction of an Hernquist model is therefore
\begin{equation}
    \label{eq:bulgecircvel}
    v_{\rm circ,b}(r) = \frac{\sqrt{G M_{\rm b} r}}{r+R_{\rm b}}.
\end{equation}

The radial velocity dispersion can be determined from the Jeans equation as 
\begin{equation}
    \label{eq:bulge_sigma_r}
    \begin{split}
    \overline{v_r^2} = & \frac{GM_{\rm b}}{12R_{\rm b}}\Biggl\{ \frac{12r(r+R_{\rm b})^3}{R_{\rm b}^4}\log{\left(\frac{r+R_{\rm b}}{r}\right)} \\ & - \frac{r}{r+R_{\rm b}}\left[25+52\frac{r}{R_{\rm b}}+42\left(\frac{r}{R_{\rm b}}\right)^2+12\left(\frac{r}{R_{\rm b}}\right)^3\right]\Biggr\}.\end{split}
\end{equation}

From the 3-D properties of the bulge, we can derive the corresponding projected quantities, namely 
\begin{itemize}
    \item the surface density
\begin{equation}
    \Sigma_{\rm b}(R) = 2 \int_R^\infty \frac{\rho(r)r}{\sqrt{r^2-R^2} }dr = \frac{M_{\rm b}}{2\upi R_{\rm b}^2 (1-s^2)^2}[(2+s^2)X(s)-3],
    \label{eq:Hernquist_projected_density}
\end{equation}
where $s=R/R_{\rm b}$, and the auxiliary variable $X$ is defined as
\begin{equation}
X(s) = \begin{cases} \frac{1}{\sqrt{1-s^2}} \mathrm{sech}^{-1}{s} = \frac{\log{\left[1+\sqrt{1-s^2}/s\right]}}{\sqrt{1-s^2}} & \ \ \ \ \ \mathrm{if} \ \ \ \ 0\leq s \leq 1 \\ \\ \frac{1}{\sqrt{s^2-1}}\sec^{-1}{s} = \frac{\cos^{-1}{\left(1/s\right)}}{\sqrt{s^2-1}} & \ \ \ \ \  \mathrm{if} \ \ \ \ s >1 \end{cases}
\label{eq:X(s)}
\end{equation} 
\item and the line of sight velocity dispersion 
\begin{equation}
\begin{split}
    \sigma_{\rm b}^2(R) &= \frac{2}{\Sigma_{\rm b}(R)}\int_R^\infty \frac{\rho \Bar{v^2_r} r }{\sqrt{r^2-R^2}} dr \\ &=  \frac{G M_{\rm b}^2}{12\upi R_{\rm b}^3 \Sigma_{\rm b}(R)} \Biggl\{ \frac{1}{2(1-s^2)^3} [-3s^2 X(s) \\& (8s^6-28s^4 +35s^2-20)  -24s^6 +68s^4 -65s^2 + 6] - 6\upi s \Biggr\}.\end{split}
    \label{eq:Hernquist_projected_dispersion}
\end{equation}
\end{itemize}
Note that the velocity distribution of the bulge is computed taking into account its own potential only (an approximation that best represents compact bulges), on the contrary, as shown in the following paragraph, the velocity distribution of the disc components takes in consideration the whole potential.

We adopt an Hernquist profile also for the dark matter halo, which is the only dark component in our model hence only contributes to the overall potential. Collisionless simulations in a $\Lambda \rm CDM$ Universe found that dark matter halos can be well described by the Navarro-Frenk-White profile \citep{Navarro1996} which is almost equivalent to the Hernquist model until their scale radius, typically larger then the resolved kinematic. 
In the following, all the quantities associated with the halo will be indicated by the subscript $_h$.

\subsection{Intrinsic parameters: Inner \& outer disc }
\label{sec:inner_outer_disc}
The other visible components of our model are two exponential thin discs. The exponential profile is a commonly adopted choice to describe the external regions of disc-like galaxies \citep{DeVaucouleurs1959} and, in the assumption of zero thickness, the circular velocity and the potential in the equatorial plane can be computed with simple equations \citep{Freeman1970}. Even though this last hypothesis may not hold in all cases, it is necessary in order to keep the model analytical and hence computationally fast. The two discs differ by total mass ($M_{{\rm d},j}$, with $j=1,2$ discriminating between the two discs), scale radius ($R_{{\rm d},j}$) and mass-to-light ratio ($\left(M/L\right)_{{\rm d},j}$), and the discussion below applies to both of them. We conventionally decided to refer to the inner disc with the subscript $j = 1$, and to the disc with the largest scale radius with the subscript $j = 2$.

The intrinsic (i.e. non-projected) surface density of the $j$-th disc is: 

\begin{equation}
    \Sigma_{{\rm int, d},j}(r) = \frac{M_{{\rm d},j}}{2 \upi R_{{\rm d},j}^2} \exp(-r/R_{{\rm d},j}),
    \label{eq:disc_density}
\end{equation}
and, since the disc is assumed to be razor-thin, the surface density projected onto the sky plane is simply
\begin{equation}
    \label{eq:disc_projected_density}
    \Sigma_{{\rm d},j}(r) = \frac{\Sigma_{{\rm int, d},j}(r)}{\cos{i}} 
\end{equation}

The circular velocity corresponding to the potential generated by the disc distribution in the disc plane is \citep{Binney_Tremaine} 
\begin{equation}
    v_{{\rm circ,d},j}(r) = \sqrt{\frac{M_{{\rm d},j}}{2R_{{\rm d},j}^2}  y^2 \left[I_0\left(y\right)K_0\left(y\right)-I_1\left(y\right)K_1\left(y\right)\right]},
    \label{eq:exp_disc_circular_velocity}
\end{equation}
where $y=r/R_{{\rm d},j}$ while $I_0$, $K_0$, $I_1$, $K_1$ are the modified Bessel functions of first and second kind. 
We evaluate the intrinsic bulk tangential velocity (i.e. the average velocity, since the velocity dispersion is considered isotropic) following the prescription presented in \citet{Satoh1980}, where it is assumed that a fraction $k_j$ of the total kinetic energy is in ordered bulk motion and the rest goes into a velocity dispersion component. $k_j$ is a free parameter of the model ranging between $-1$ and $1$. For instance, if $k_j=1$ the disc is exactly in circular motion \footnote{\fr{$k_j=-1$ describes a disc in circular motion with counter-rotating orbits. In this work we assumed only co-rotating orbits since the galaxy analysed does not show any peculiar kinematic signatures in its velocity field.}}, while if $k_j=0$ the disc is entirely supported by dispersion. 
Under this assumption the intrinsic bulk tangential velocity of each disc is 
\begin{equation}
    v_{{\rm int, d},j}(r) = \sqrt{v_{\rm circ,b}^2(r)+v_{\rm circ,d,1}^2(r)+v_{\rm circ,d,2}^2(r)+v_{\rm circ,h}^2(r)},
\end{equation}
where $v_{\rm circ,h}$ is the circular velocity associated with the dark matter halo potential computed as in eq. \ref{eq:bulgecircvel} using the total mass and scale radius of the halo. The bulk l.o.s. velocity component at any location in the plane of the sky $\mathbfit{R}$ is then
\begin{equation}
    \label{eq:disc_projected_velocity}
    v_{{\rm d},j}(\mathbfit{R}) = -k_j\, \sin{i}\, \cos{\phi}\, v_{{\rm int, d},j}(r),
\end{equation}
where $r$ is the deprojected distance from the centre of the galaxy of the $\mathbfit{R}$ position and $\phi$ is the azimuthal angle evaluated in the plane of the disc from the major axis of the projected disc.

The l.o.s. velocity dispersion of the disc is instead
\begin{equation}
    \label{eq:disc_projected_dispersion}
    \sigma^2_{{\rm d},j}(\mathbfit{R}) = (1-k_j^2)\frac{v^2_{{\rm int, d},j}(r)}{3},
\end{equation}
where the factor 3 at the denominator is due to the assumption of an isotropic velocity dispersion tensor. \fr{Note that in our model the discs are thought to have a finite, even though small, thickness and the razor-thin disc assumption is only an approximation of their description. This implies that the vertical velocity dispersion is not zero, justifying the $1/3$ term in eq.~\ref{eq:disc_projected_dispersion}.}

As we will see in the following analysis, the introduction of the inner disc component, which is partially superimposed to the bulge, can affect the kinematics increasing the rotational support in the central regions. The system "bulge+inner disc" can be considered as an approximate description for a rotating (pseudo)-bulge.

\subsection{Full model observables}
\label{sec:total_quantities}. 

In general, since our model is the superposition of different potential-density pairs, total quantities are computed through \fr{light-}weighted averages of each component. 
 
Using Eqs. (\ref{eq:Hernquist_projected_density},\ref{eq:disc_projected_density}) for each position $\mathbfit{R}$ in the sky, we determine the total surface brightness as
\begin{equation}
    \label{eq:total_brightness}
    \begin{split}
    B_{tot}(\mathbfit{R}) &= B_{\rm b}(\mathbfit{R})+B_{{\rm d},1}(\mathbfit{R})+B_{{\rm d},2}(\mathbfit{R}) \\ &=\left(\frac{M}{L}\right)^{-1}_{\rm b} \Sigma_{\rm b}(\mathbfit{R}) + \left(\frac{M}{L}\right)^{-1}_{{\rm d},1} \Sigma_{{\rm d},1}(\mathbfit{R}) + \left(\frac{M}{L}\right)^{-1}_{{\rm d},2} \Sigma_{{\rm d},2}(\mathbfit{R}),
    \end{split}
\end{equation}

where $\left(M/L\right)_{\rm b}$, $\left(M/L\right)_{{\rm d},1}$ and $\left(M/L\right)_{{\rm d},2}$ are the constant mass to light ratios of the bulge, the inner disc and the outer disc. 

The line of sight velocity, as anticipated, is the average weighted on each surface brightness of the velocities of the three components:
\begin{equation}
    \label{eq:total_velocity}
    v_{\rm los}(\mathbfit{R}) = \frac{B_{{\rm d},1}(\mathbfit{R})v_{{\rm d},1}(\mathbfit{R})+B_{{\rm d},2}(\mathbfit{R})v_{{\rm d},2}(\mathbfit{R})}{B_{\rm b}(\mathbfit{R})+B_{{\rm d},1}(\mathbfit{R})+B_{{\rm d},2}(\mathbfit{R})},
\end{equation}
where no bulge contribution is included in the numerator due to its isotropic velocity field.

Similarly, the projected dispersion can be computed as
\begin{equation}
    \label{eq:total_dispersion}
    \sigma_{\rm los}^2(\mathbfit{R}) = \frac{B_{\rm b}(\mathbfit{R})\sigma_{\rm b}^2(\mathbfit{R})+B_{{\rm d},1}(\mathbfit{R})\sigma_{{\rm d},1}^2(\mathbfit{R})+B_{{\rm d},2}(\mathbfit{R})\sigma_{{\rm d},2}^2(\mathbfit{R})}{B_{\rm b}(\mathbfit{R})+B_{{\rm d},1}(\mathbfit{R})+B_{{\rm d},2}(\mathbfit{R})},
\end{equation}
where $\sigma_{\rm b}$ depends only on the magnitude of $\mathbfit{R}$, due to the spherical symmetry of the bulge model \fr{and the average is weighted by the contribution of each component to the total luminosity.}

The 17 model parameters used in the reference model are summarized in table \ref{tab:parameters}.

\begin{table}
	\centering
	\caption{Summary of the model parameters.}
	\label{tab:parameters}
	\begin{tabular}{lr} 
		\hline
		name & description \\
		\hline
		$x_0$ ; $y_0$ & horizontal/vertical position of the center\\
		$\mathrm{P.A.}$ & position angle \\
		$i$ & inclination angle \\
		$M_{\rm b}$ ; $R_{\rm b}$ & mass and scale radius of the bulge \\
		$M_{{\rm d},1}$ ; $R_{{\rm d},1}$ & mass and scale radius of the inner disc \\
		$M_{{\rm d},2}$ ; $R_{{\rm d},2}$ & mass and scale radius of the outer disc \\
		$M_{h}$ ; $R_h$ & mass and scale radius of the halo \\
		$(M/L)_{\rm b}$ & mass-to-light ratio of the bulge \\
		$(M/L)_{{\rm d},1}$ & mass-to-light ratio of the inner disc \\
		$(M/L)_{{\rm d},2}$ & mass-to-light ratio of the outer disc \\
		$k_1$ & kinematical decomposition parameter of the inner disc \\
		$k_2$ & kinematical decomposition parameter of the outer disc \\
		\hline
	\end{tabular}
\end{table}

\subsection{Parameter Estimation}
\label{sec:parameter_estimation}

In the presence of such a large parameter space standard fitting routines, such as $\chi^2$ minimization by gradient descend, often fail. This is mainly due to the presence of multiple local minima and strong correlations between the parameters. In these situations a possible solution can be found in genetic algorithms, such as particle swarm optimization \citep{PSO}, which directly search in the parameter space using a momentum based approach, without considering any gradient. 
Even though these methods are quite fast, since they require only few likelihood evaluations,  they usually depend on a number of hyperparameters which must be carefully tuned in order to ensure convergence to the correct maximum. Moreover, we are particularly interested in exploring the parameter space in its entirety, thus to be able to map the full $n$-dimensional probability distribution, including all correlations among parameters. Furthermore, we wish to be able to compare the predictions of several models and select among them the ones that better reproduce the observations.
For these reasons, we decided to operate within the framework of Bayesian inference and to adopt the nested sampling Bayesian algorithm to reconstruct the full probability distribution for all parameters of interest as well as for the estimation of the evidence. More specifically, we use the \textsc{python} package CPNest\footnote{\url{https://github.com/johnveitch/cpnest}} \citep{Del_Pozzo_CPNest}, which allows for multi-threading and offers different sampling techniques.\footnote{In our specific implementation we have selected a slice sampler \citep{Neal_2003}.}

We will now briefly summarize the most important features of the nested sampling algorithm in order to understand when it might be preferable to a standard Markov Chain Monte Carlo (MCMC).

Bayes theorem states that
\begin{equation}
    \label{eq:Bayes_theorem}
    p({\Omega}|data) = \frac{\mathcal{L}(\textit{data}|{\Omega}) \times \pi({\Omega})}{Z(\textit{data})},
\end{equation}

where ${\Omega}$ is the parameter vector, $p$ is the posterior probability distribution, $\mathcal{L}$ is the likelihood function, $\pi$ is the prior distribution and $Z$ is the evidence. Usually, MCMC methods draw a fixed number of samples directly from the posterior probability making use of the fact that it is proportional to the product between the likelihood and the prior. $Z$ is considered as a normalization factor, hence in the exploration step it simplifies.

The primary target of the nested sampling method is, instead, the evidence itself. The algorithm starts by sampling $N$ parameters vectors ${\Omega}_{i=1,..,N}$, usually referred as "live points", from the prior probability. Then all the live points are sorted by their likelihood such that $\mathcal{L}_{1}<...<\mathcal{L}_N$. At each iteration $\tau$, the parameter vector associated to the lowest likelihood value is replaced by a new live point ${\Omega}_k$ such that $\mathcal{L}_k>\mathcal{L}_1$\footnote{Note that finding ${\Omega}_k$ is not trivial and it is usually done with standard MCMC sampling techniques. Note also that we only need to identify the worst live point (i.e. the one with the lowest likelihood) without sorting the other $N-1$.}. Given the live point with the lowest likelihood ($\mathcal{L}_1$ in this case) we can define the enclosed prior mass $X_1$ as the integral of the prior over all the parameters $\Omega_i$ such that $\mathcal{L}_i>\mathcal{L}_1$\footnote{Note that with such a definition $0\leq X\leq1$ is a monotonically decreasing function of $\mathcal{L}$.}. Now, starting from $X_0=1$ we can compute the prior mass at iteration $\tau=1$ as $X_\tau=t_\tau X_{\tau-1}$ where $t_\tau$ is the largest of N random numbers uniformly drawn between 0 and 1 and increment the evidence by $\sim \mathcal{L_{\tau}}(X_{\tau-1}-X_{\tau})$. 
This operation is repeated until further increases in $Z$ will not change its value more that a small fraction $f$.

The existence of a stopping criterion based on the accuracy of the evidence calculation is a major difference compared to standard MCMC methods where the number of iterations must be chosen \textit{a priori} and, especially in cases with complicated posteriors, it is often unclear how large it should be. The evidence is critical for model selection, since, given two competing models $H_1$ and $H_2$, one can compute the \emph{odds ratio} as:
\begin{equation}
    O_{12} = \frac{p(H_1)}{p(H_2)}\frac{p(D|H_1)}{p(D|H_2)} =  \frac{p(H_1)}{p(H_2)}\frac{Z_1}{Z_2}\equiv\frac{p(H_1)}{p(H_2)}B_{12}
\end{equation}
where $p(H_1)/p(H_2)$ is the prior odds that can be typically set to 1, and $Z_1/Z_2$,the ratio of the model evidences obtained by integrating over the entire parameter spaces for $H_1$ and $H_2$, is called the Bayes factor. 
If $B_{12}$ is > 1 (< 1) then model $H_1$ is favoured (disfavoured) compared to $H_2$ (\fr{see \citealt{Jeffreys_probability} for a detailed classification}). The Bayes factor offers a practical way of discriminating among models by taking into account the entirety of the posterior distribution for the model parameters.

\subsubsection{The likelihood}
\label{sec:the_likelihood}
From Eq.~\eqref{eq:Bayes_theorem} we can see that the likelihood function $\mathcal{L}$ plays a fundamental role in Bayesian inference. The likelihood, in fact, quantifies how likely the data are given a specified model, hence it quantifies the predictive power of a model. In our case the likelihood is composed by 4 different terms, each of them modelled assuming a Gaussian distribution, and reads

\begin{equation}
    \label{eq:likelihood}
    \log{\mathcal{L}} = \log{\mathcal{L}_{B}}+\log{\mathcal{L}_{v_{los}}}+\log{\mathcal{L}_{\sigma_{los}}}+\log{\mathcal{L}_{M/L}}.
\end{equation}

The first three terms compare the output of our model (i.e. Eq \ref{eq:total_brightness}, \ref{eq:total_velocity}, and \ref{eq:total_dispersion}) with the data, while the last one contains additional information regarding the $M/L$ ratio.

From our modelling we can easily compute an average mass-to-light ratio as

\begin{equation}
    \langle L/M\rangle=\frac{\left(\frac{M}{L}\right)^{-1}_{\rm b}\Sigma_b +\left(\frac{M}{L}\right)^{-1}_{\rm d,1}\Sigma_{d,1} +\left(\frac{M}{L}\right)^{-1}_{\rm d,2}\Sigma_{d,2}}{\Sigma_b+\Sigma_{d,1}+\Sigma_{d,2}},
    \label{eq:ML_ratio}
\end{equation}

\footnote{We thus approximate the average mass-to-light ratio as $\langle M/L \rangle = 1/ \langle L/M \rangle$ } Since we work with IFU observations it is possible to compute the two dimensional $M/L$ ratio maps and its error comparing the spatially resolved spectra with stellar population synthesis (SPS) models (for a more complete description see \citealt{Zibetti_2017}).

In the case that an SPS modelling is not doable we provide our fitting tool with a preprocessing routine that compute the resolved $M/L$ ratio based on multiple band photometry and color--$M/L$ relations \citep{Garcia_2018,Zibetti_2009}. More precisely the $M/L$ ratio is computed as

\begin{equation}
    \label{eq:ML-color}
    \log_{10}{(M/L)_{\alpha}}= a_{\alpha}+b_{\alpha}\times(m_{\beta}-m_{\gamma})
\end{equation}
where $m_k$ are the magnitudes in the $k=\alpha$, $\beta$ and $\gamma$ bands used to estimate $M/L$ and ($a_{\alpha}$, $b_{\alpha}$) are two constants calibrated against different independent estimates of the ratio. Note that this relation can use three different bands or assume either $\alpha=\beta$ or $\alpha=\gamma$.

In the following analysis of the NGC 7683 galaxy we used spatially resolved $M/L$ ratio computed from SPS modelling. We refer to Appendix~\ref{app:ML_from_color} for more details about the analysis with multiple bands photometry and for a comparison between the two methods. 

Although the evaluation of the likelihood on a single central processing unit (CPU) has a moderate computational cost, i.e. fractions of a second, the nested sampling algorithm requires a huge amount of likelihood evaluations, which usually results in days or event weeks of computation for a single run. This major drawback motivated us to move part of the algorithm on GPUs using the \textsc{python} package \textsc{numba}. We reached so far a total duration for the parameter estimation of about 3 hours with a boost of more than a factor of 100 on an NVIDIA TESLA V100.

\section{Results}
In the following paragraphs, we will show the results of the application of out fitting tool to the test galaxy NGC 7683, and compare them with those obtained using well-tested independent methods.

\subsection{NGC 7683}
\label{sec:our_analysis}

\fr{NGC 7683 is an S0 galaxy at redshift $z=0.01227$\ ($\sim 53\,\textrm{Mpc}$) with a mild inclination and a regular velocity pattern that can be considered a representative example of the disc galaxies population in the local Universe. Even though our assumptions of razor-thin disc with a clear bulge to disc decomposition may fit better Sa-Sb galaxies we instead opted for an S0 galaxy. Lenticular galaxies are usually dominated by old stellar populations, affected by little or negligible dust attenuation, thus resulting in homogeneous M/L ratios and prominent absorption features in their spectra, which facilitate the extraction of the kinematics. Among the S0s in the CALIFA survey we selected NGC 7683 by visually inspecting its highly regular photometry and kinematics (no isophotal distortions, rings, spiral structures, counter rotation etc.) resulting in an optimal candidate for a first test of our implementation.}
NGC 7683 has been observed in IFS as part of the main sample of the CALIFA survey \citep{Sanchez_2012}. Our work is based on the results of the analysis performed by \cite{Zibetti_2017}, who joined the information from the CALIFA Data Release 3 \citep[DR3][]{Sanchez_2016} ``COMBO'' cubes and from the Sloan Digital Sky Survey \citep{SDSS} imaging in the \textit{ugriz} bands. The fully detailed description of the analysis can be found in \cite{Zibetti_2017}, here we recall the features that are most relevant to the present analysis. The adopted ``COMBO'' cube from CALIFA provides a spectral coverage from 3700 to 7140 \AA~ at the resolution of $\sim 120\, \rm km/s$, which results from the joining of the two observing setups (red, low-resolution and blue, high-resolution) of the survey. The native spatial resolution is $\sim 2.6\arcsec$, but the cube is smoothed in the 2D spatial plane with an adaptive kernel, using the IFS version of the ADAPTSMOOTH code \citep[see][]{Zibetti_2009,adaptsmooth}. This procedure provides the best trade-off between spectral signal-to-noise ratio (sufficient for reliable stellar population and kinematic analysis) and spatial resolution, by preserving the maximum possible information about the galaxy structure. The cube has been analysed using the PPXF \citep{cappellari_ensellem_2004} and GANDALF \citep{sarzi_2006} in order to derive accurate kinematic information (first and second moment of the l.o.s. velocity distribution, $v_{sys}$ and $\sigma$) and to subtract the ionized-gas line emission from the stellar continuum.
The images in the five \textit{ugriz} bands are taken from the nineth data release of the SDSS \citep[DR9][]{Ahn+2012} and are matched to the spatial resolution of the smoothed CALIFA cubes as far as stellar population analysis is concerned. Using the bayesian stellar population inference method described in \cite{Zibetti_2017} we derive maps of various stellar population parameters, including the stellar mass-to-light ratio, which will be used next in this work.

\fr{Since the errors from the raw image cannot be easily retrieved and propagated during the reduction and calibration process, we assumed for the surface brightness $B$ a standard 5\% error, comparable with the relative error in the kinematic quantities. We added a constant systematic error in the surface brightness too, that we estimate to be $\simeq 10 \%$ of its minimum value (i.e., $\simeq 5\rm L_{\sun}/pc^2 $). This implies that the relative error on the surface brightness increases moving outwards. Note that the systematic error in $B$ could be treated as an additional fitting parameter in the likelihood, rather than fixed a priori as done in this work. Given that, as we verified, the two methods actually give similar results, we opted to reduce the number of fitting parameters.}
The errors on the line of sight velocity and on the line of sight velocity dispersion ($\delta v$, $\delta \sigma$) directly come from the spectral analysis and are, on average, about $10\, \rm km/s$ and $15\, \rm km/s$.

Before comparing the predictions of our model to the observed data, we have to modify Eq.s~(\ref{eq:total_brightness},\ref{eq:total_velocity},\ref{eq:total_dispersion},\ref{eq:ML_ratio}) in order to account for the PSF. We thus assume a Gaussian PSF with an instrument-dependent width $\sigma$, which gives the convolution\footnote{Note that since we work on pixellated maps, the convolution is given by discrete sums.} in each point ($\tilde{x}$, $\tilde{y}$) as

\begin{equation}
    \label{eq:psf_convolution}
    F(\tilde{x},\tilde{y}) = \frac{\sum_{i,j} f(x_i,y_j)\times \exp\{-\left[(x_i-\tilde{x})^2+(y_i-\tilde{x})^2\right]/2\sigma^2\}}{\sum_{i,j}\exp\{-\left[(x_i-\tilde{x})^2+(y_i-\tilde{x})^2\right]/2\sigma^2\}},
\end{equation}
where $f$ can be $B_{tot}$, $v_{los}$, $\sigma_{los}$, or $\left<L/M\right>$, $F$ is the convolved function, and $x_i$ and $y_j$ are linearly spaced points centered at ($\tilde{x}$, $\tilde{y}$) within a radius $3\sigma$.

The best fit parameters estimated by applying our fitting tool to the NGC 7683 galaxy are reported in Tab.~\ref{tab:best_fit_parameters} together with the associated errors and the priors imposed to initialise the nested sampling. The priors are either uniform in a relatively broad range or, for the parameters possibly varying by several orders of magnitude, log-uniform, with the exception of the inclination angle $i$, that is weighted as the corresponding solid angle. Note that from our analysis it is not possible to fully gauge the halo parameters likely because of the limited spatial extension of the kinematic data, nevertheless we give an estimation of the halo mass within 5 disc scale radii ($M_{{\rm h},5}$).

\begin{table}
	\centering
	\caption{Results of the fit over the NGC 7683 galaxy. The first column is the name of the parameter, the second one refers to the uniform prior range and the third one shows the best fit value and the errors estimated from the posterior probability using the median and the $16$ and $84$ percentiles. Note that we do not have estimations for $M_{\rm h}$ and $R_{\rm h}$ due to the lack of kinematic information at large radii, but we can roughly infer the mass enclosed within $5R_{\rm d,2}$. }
	\label{tab:best_fit_parameters}
	\begin{tabular}{lcr} 
		\hline
		Parameter & Prior range & Best fit value \\
		\hline \rowsep
		$x_0 [\mathrm{kpc}] $ &  $[-2,1]$ & $-1.031^{+0.002}_{-0.002}$ \\ \rowsep
		$y_0 [\mathrm{kpc}]$ &  $[-2,1]$ & $-1.054^{+0.002}_{-0.002}$\\  \rowsep
		$\rm P.A. [\mathrm{deg}]$ &  $[-180,180]$ & $44.12^{0.06}_{0.06}$ \\ \rowsep
		$\sin{i}$ &  $[0.17,0.97]$ & $0.829^{0.001}_{0.001}$\\ \rowsep
		$\log_{10}(M_{\rm b}/\rm M_{\sun}) $ &  $[8.5;11.5]$ & $10.67^{0.01}_{0.01}$\\ \rowsep
		$\log_{10}(R_{\rm b}/\mathrm{kpc}) $ &  $[-2,1]$ & $-0.65^{0.02}_{0.02}$ \\ \rowsep
		$\log_{10}(M_{\rm d,1}/\rm M_{\sun})$ &  $[8.5;11.5]$ & $10.42^{0.02}_{0.02}$ \\  \rowsep
		$\log_{10}(R_{\rm d,1}/\mathrm{kpc}) $ &  $[-2,1]$ & $-0.015^{0.003}_{0.003}$ \\ \rowsep
		$\log_{10}(M_{\rm d,2}/\rm M_{\sun})$ &  $[8.5;11.5]$ & $10.865^{0.003}_{0.003}$ \\ \rowsep
		$\log_{10}(R_{\rm d,2}/\mathrm{kpc}) $ &  $[-2,1]$ & $0.603^{0.002}_{0.002}$ \\ \rowsep
		$\log_{10}(M_{\rm h}/\rm M_{\sun})$ &   $[11.5,14.5]$ & - \\ \rowsep
		$\log_{10}(R_{\rm h}/\mathrm{kpc})$ &  $[1;3]$ & - \\ \rowsep
		$\log_{10}{M_{\rm h,5}/\rm M_{\sun}}$ & - & $11.862^{0.006}_{0.006}$ \\ \rowsep
		$(M/L)_{\rm b} [\rm{M_{\sun}/L_{\sun}}]$ &  $[1;100]$ & $10.5^{0.2}_{0.2}$ \\ \rowsep
		$(M/L)_{\rm d,1} [\mathrm{M_{\sun}/L_{\sun}}]$ &  $[0.1;100]$ & $2.12^{0.06}_{0.06}$ \\ \rowsep
		$(M/L)_{\rm d,2} [\mathrm{M_{\sun}/L_{\sun}}]$ &  $[0.1;100]$ & $2.62^{0.02}_{0.02}$ \\ \rowsep
		$k_1$ &  $[0;1]$ & $0.299^{0.002}_{0.002}$ \\ \rowsep
		$k_2$ &  $[0;1]$ & $0.700^{0.002}_{0.001}$ \\
		\hline
	\end{tabular}
\end{table}

\begin{figure*}
    \centering
    \includegraphics[scale=0.093]{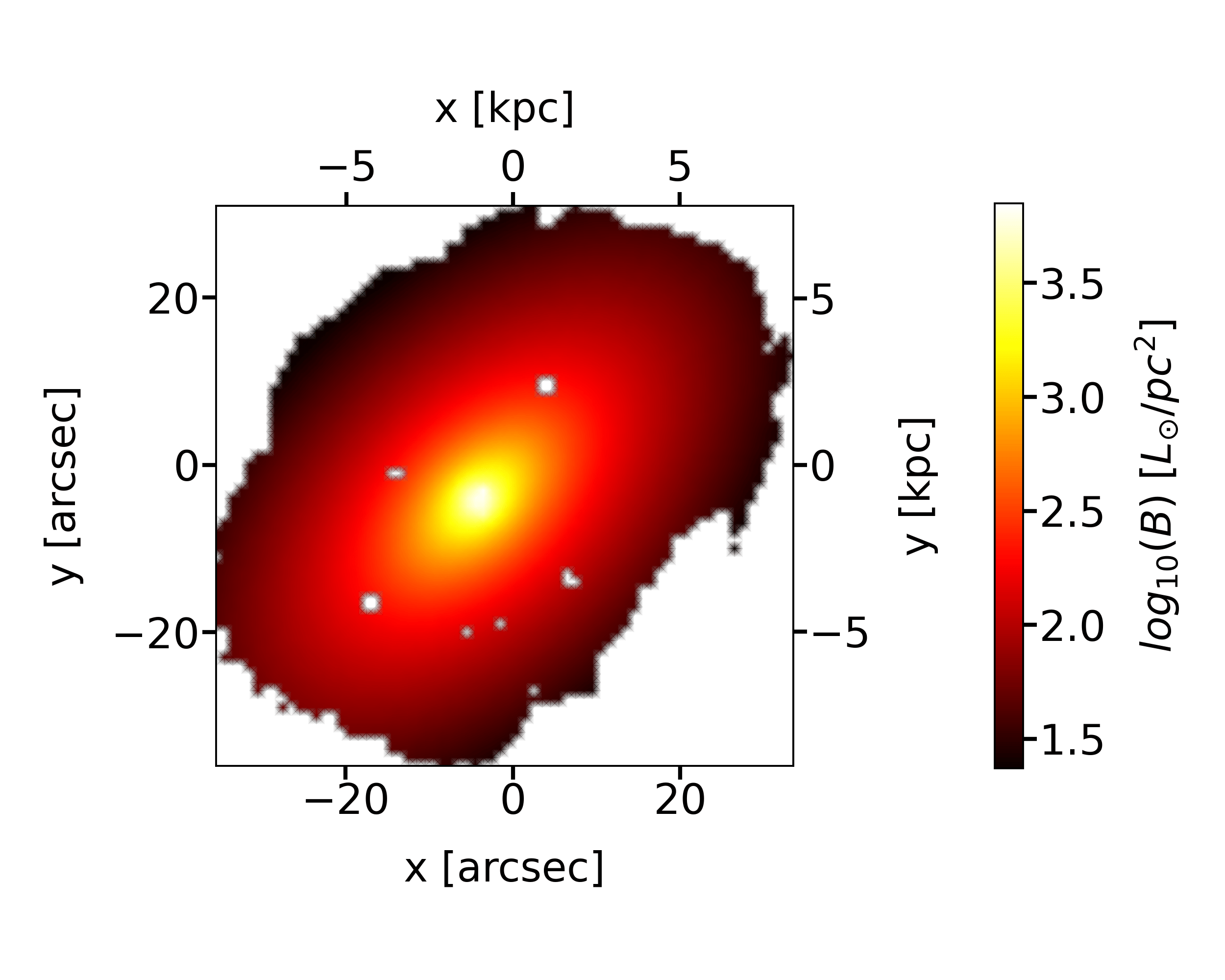}
    \includegraphics[scale=0.093]{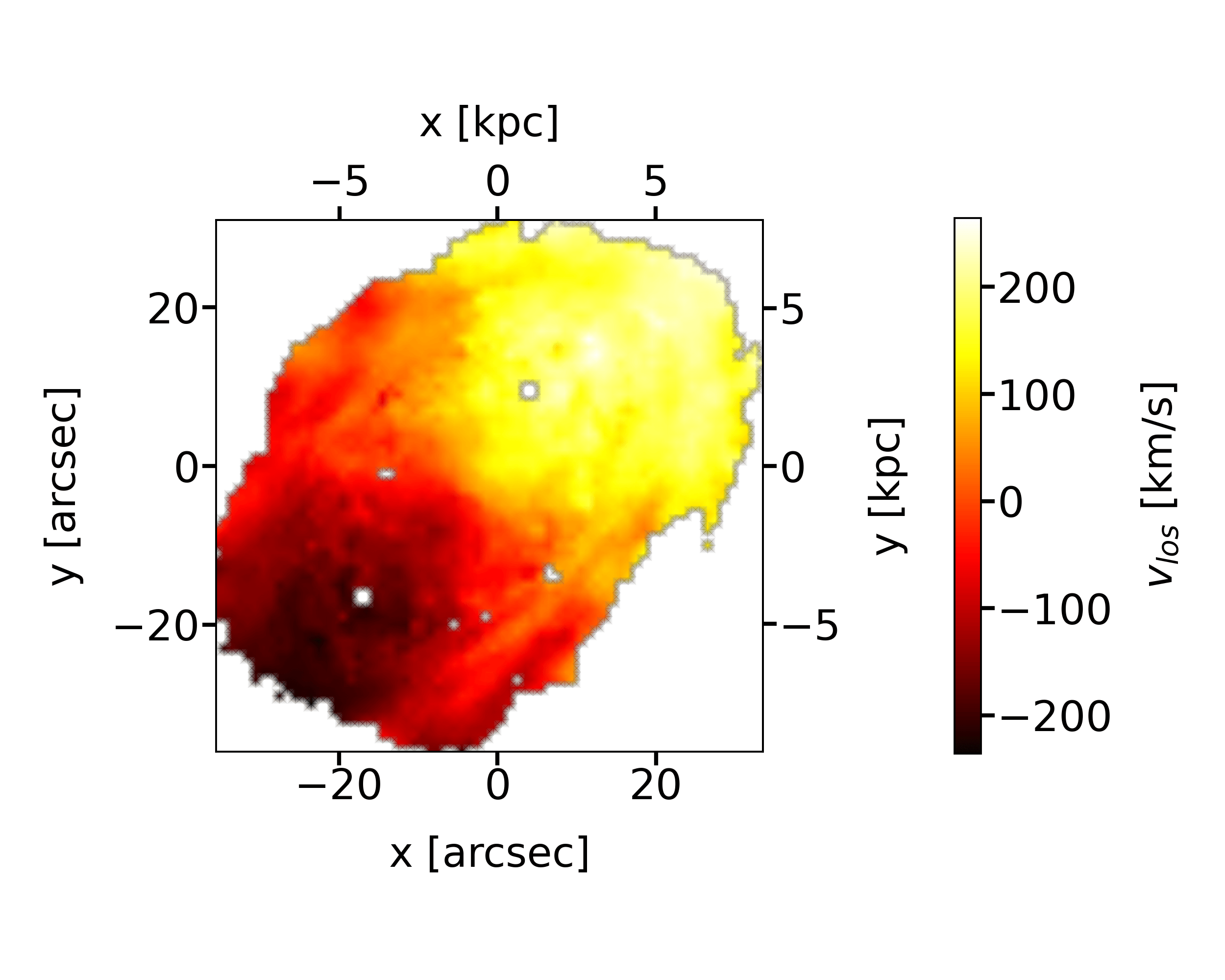}
    \includegraphics[scale=0.093]{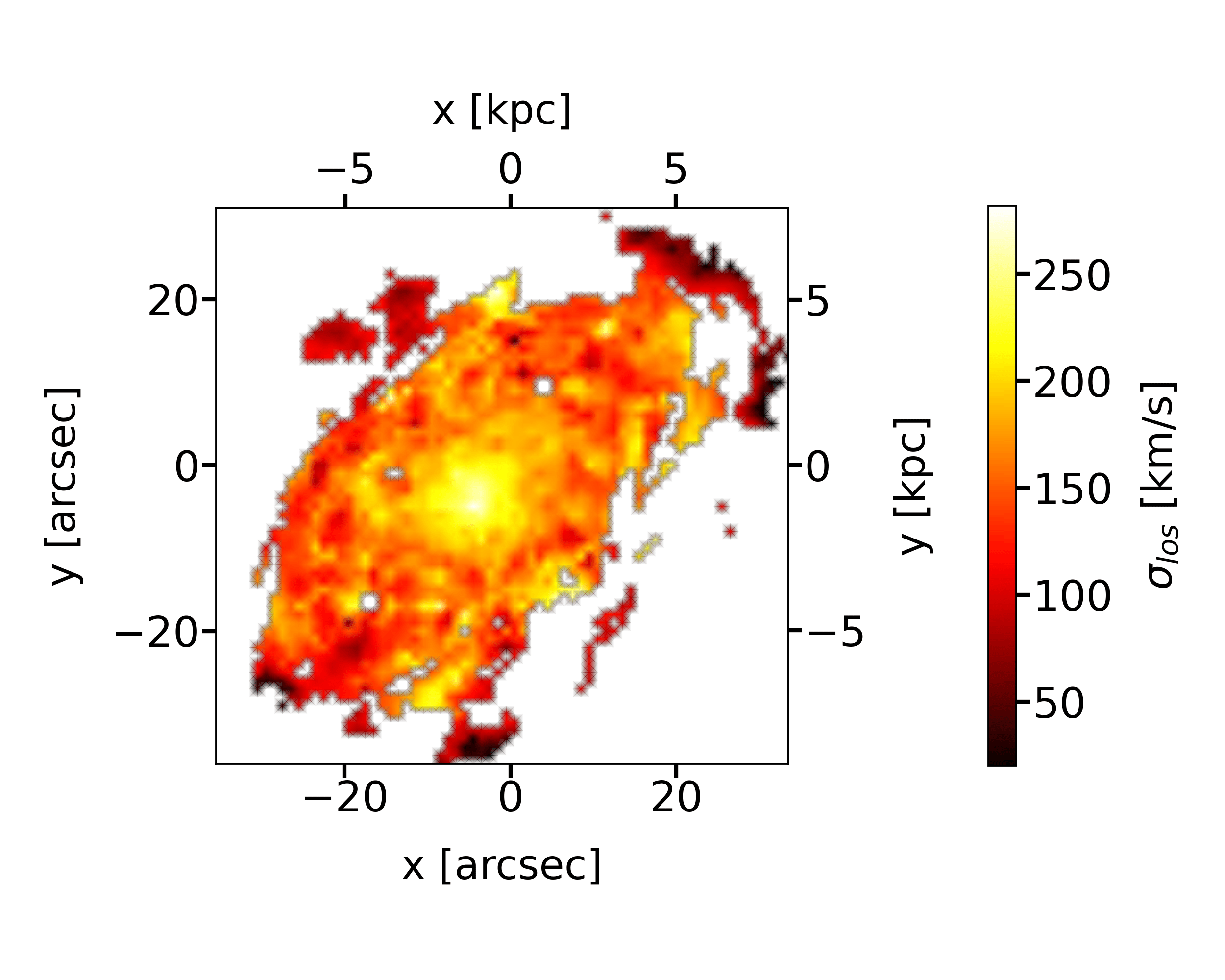}
    \includegraphics[scale=0.093]{pictures/run103/B_model_2D.png}
    \includegraphics[scale=0.093]{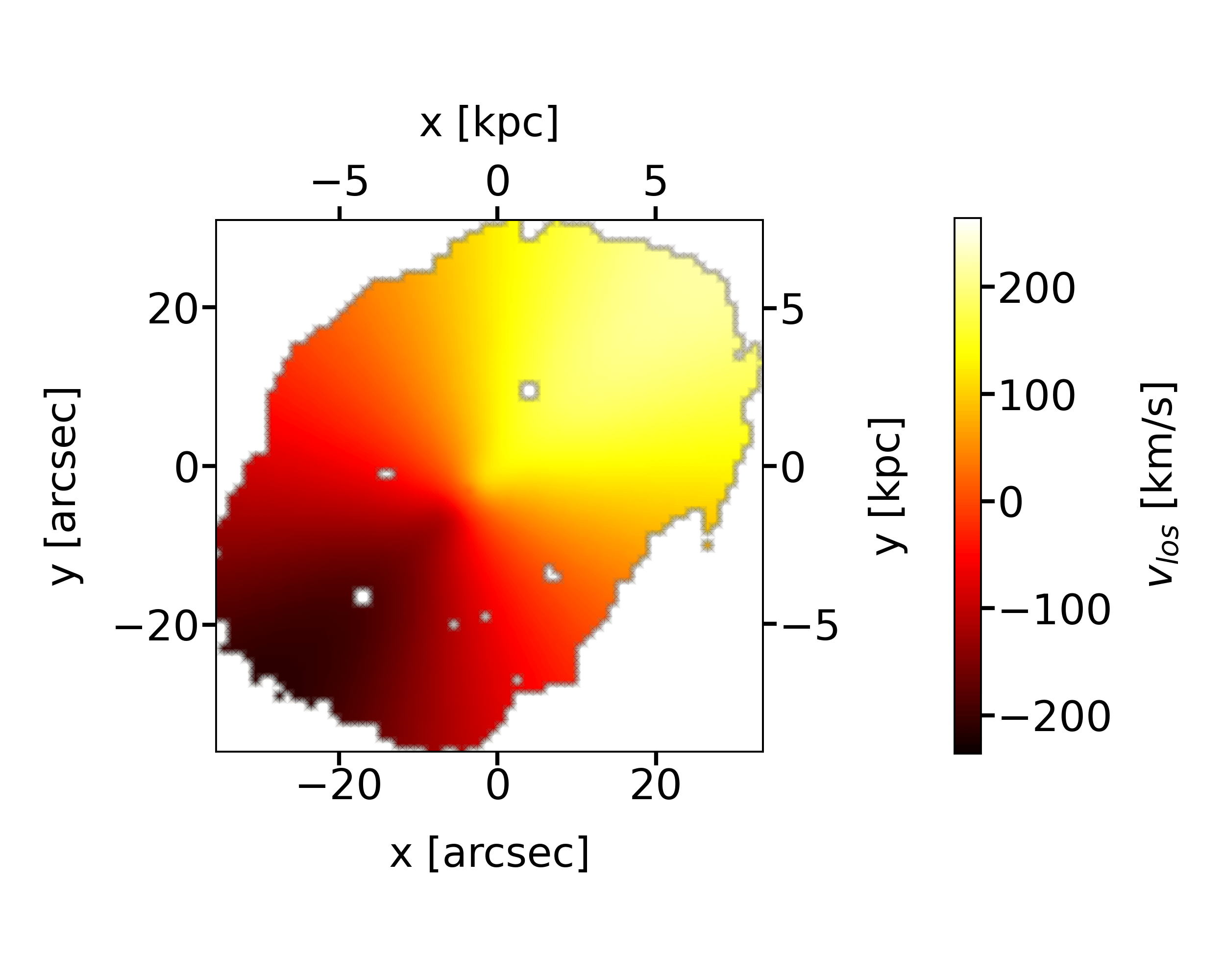}
    \includegraphics[scale=0.093]{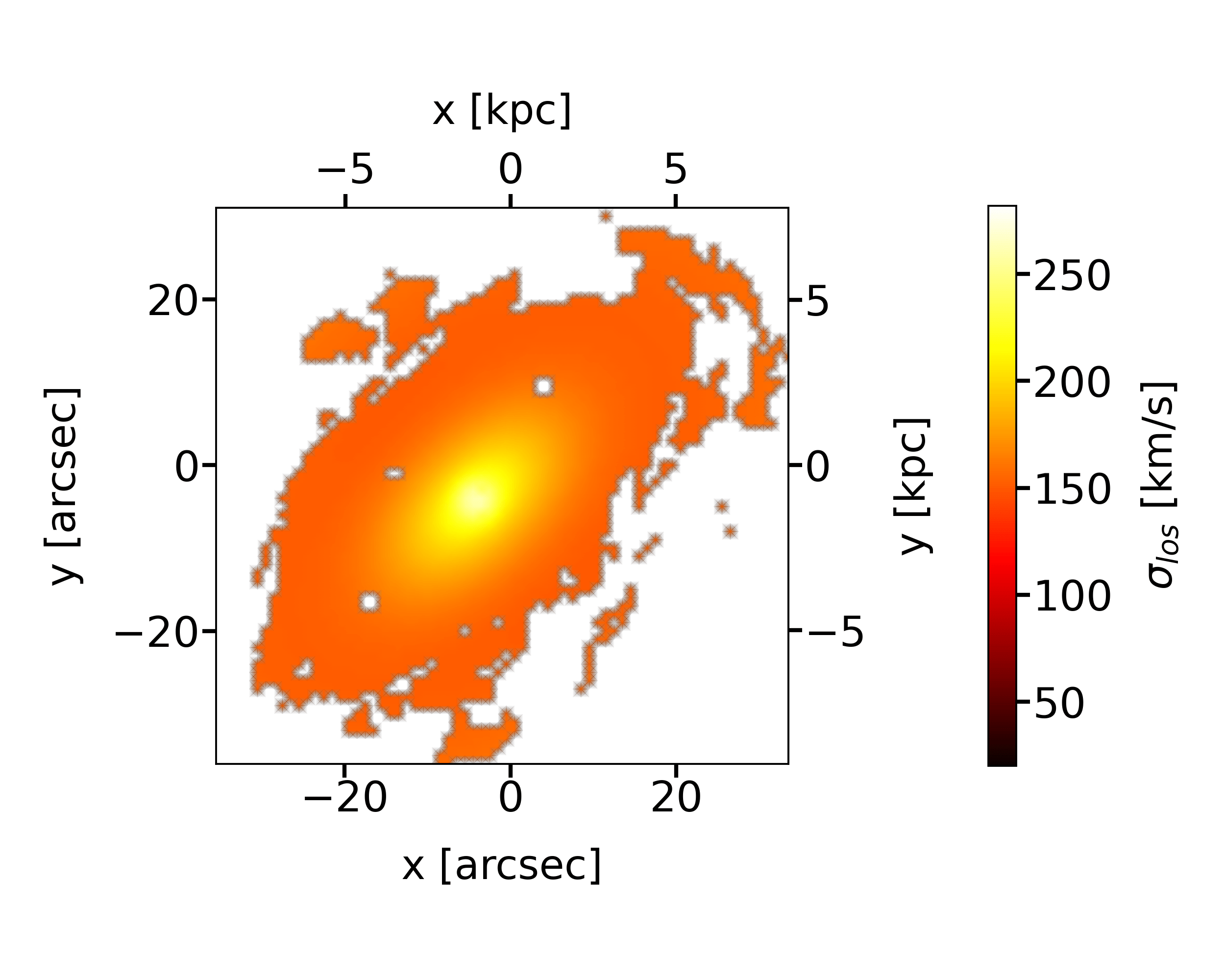}
    \includegraphics[scale=0.093]{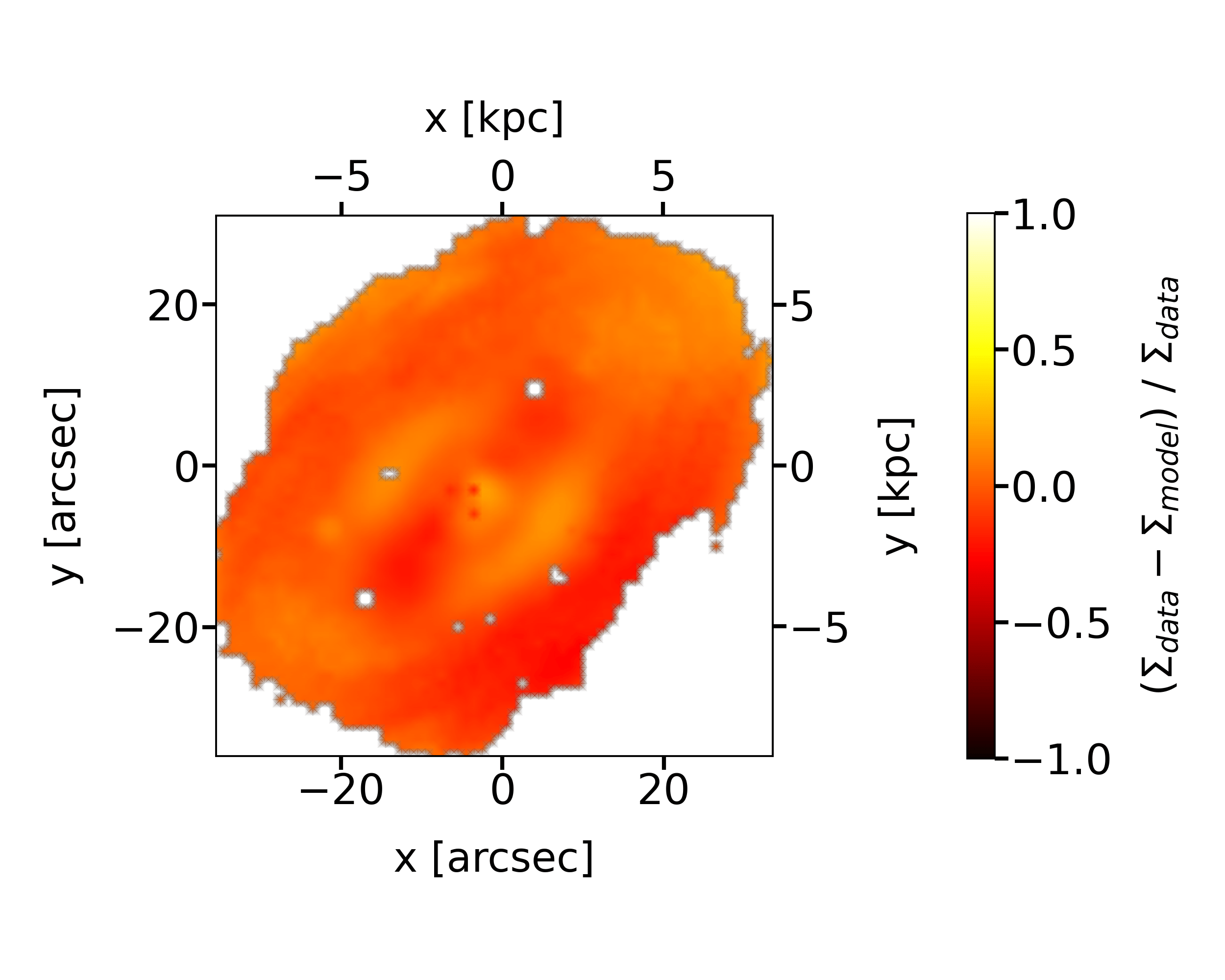}
    \includegraphics[scale=0.093]{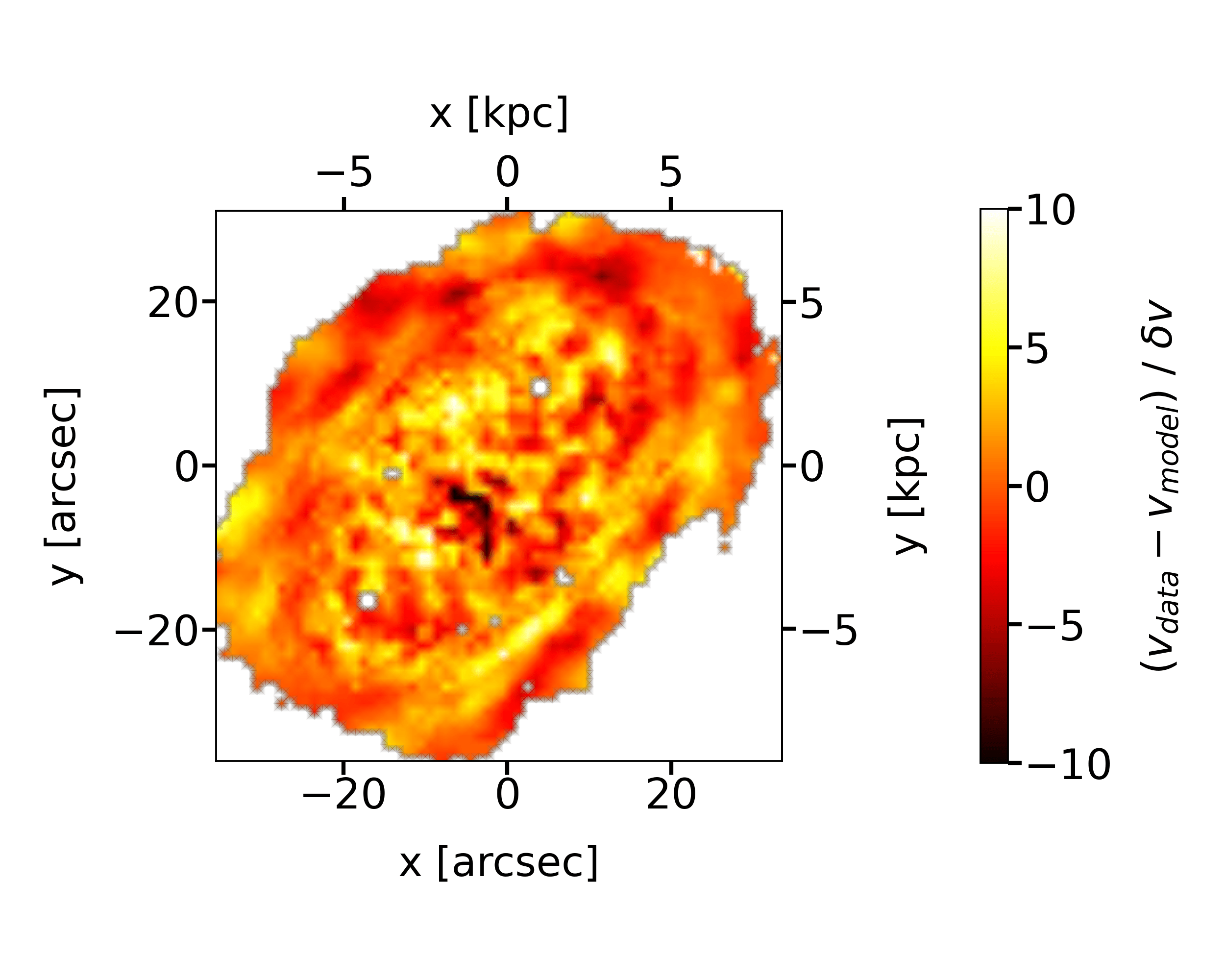}
    \includegraphics[scale=0.093]{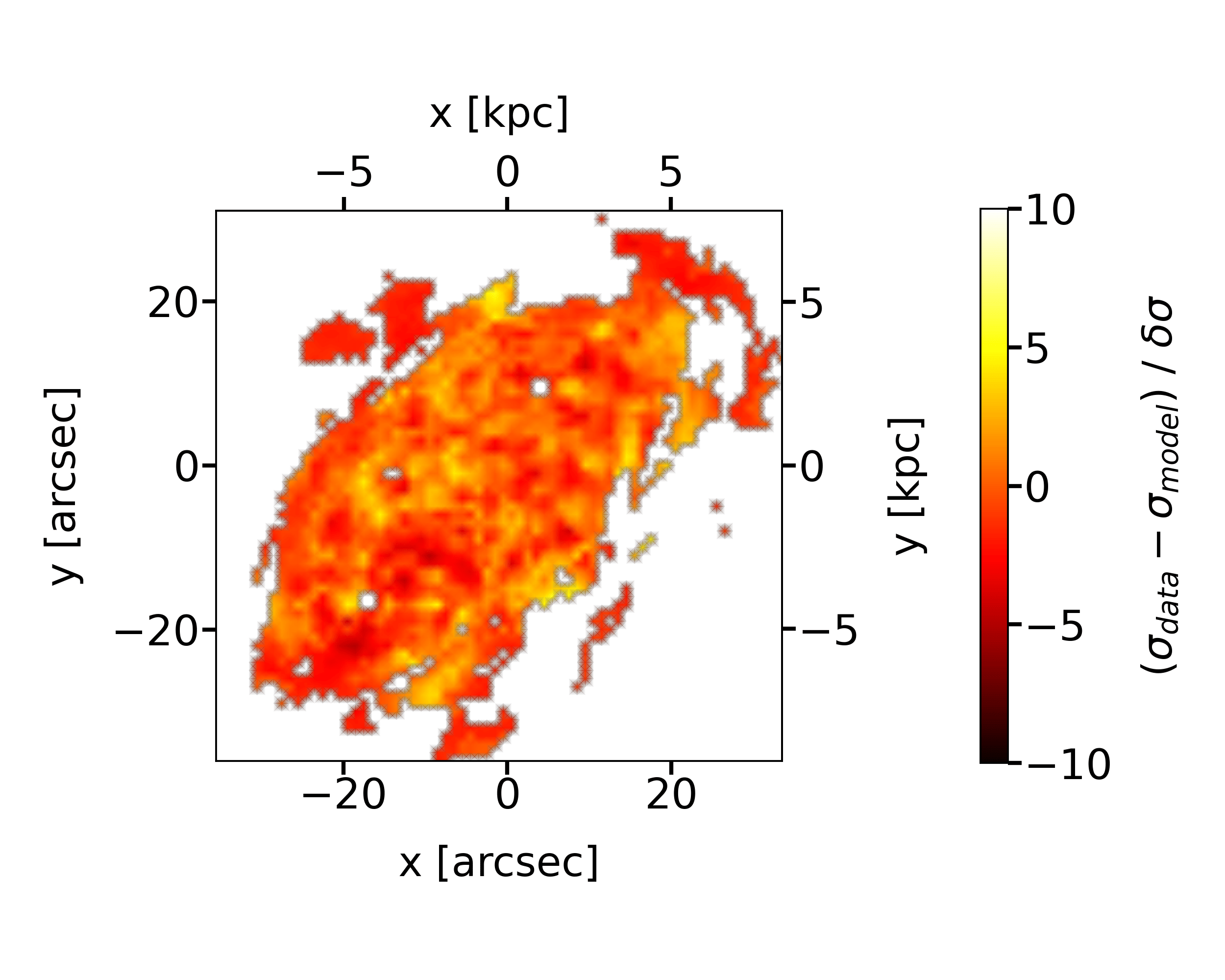}\\
    \hspace{-0.075\textwidth}
    \includegraphics[width=0.23\textwidth,height=0.18\textheight]{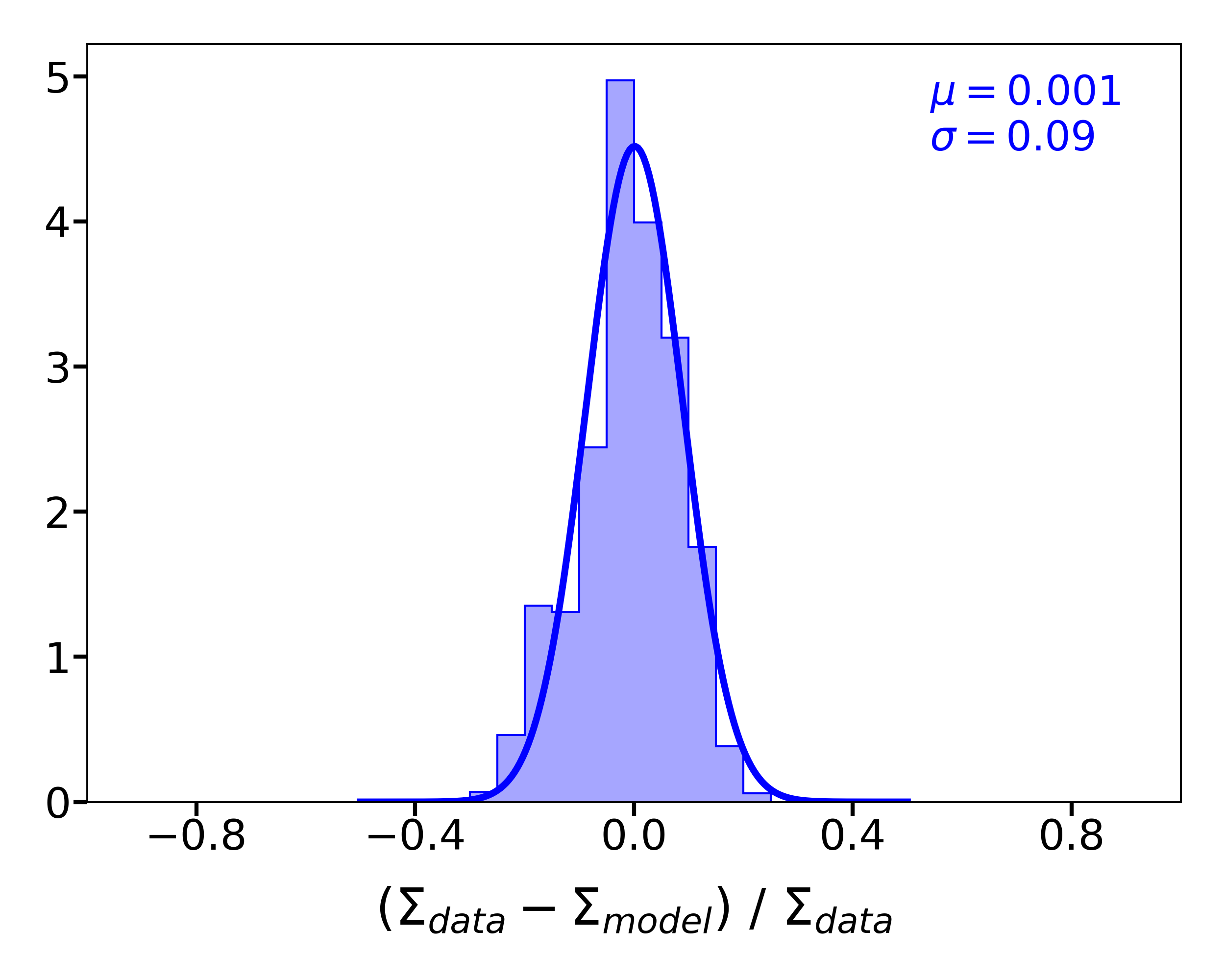}\hspace{0.09\textwidth}
    \includegraphics[width=0.23\textwidth,height=0.18\textheight]{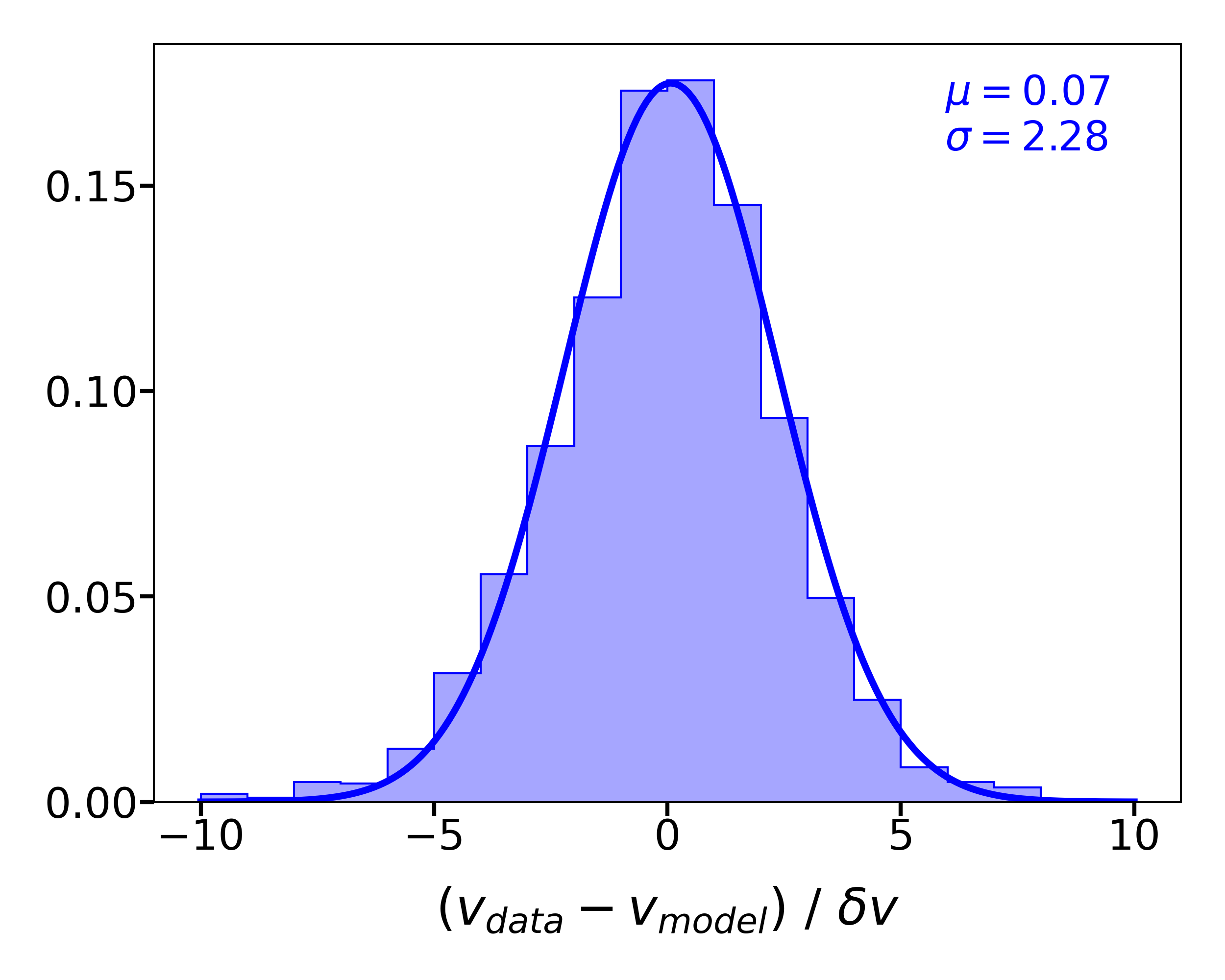}\hspace{0.105\textwidth}
    \includegraphics[width=0.23\textwidth,height=0.18\textheight]{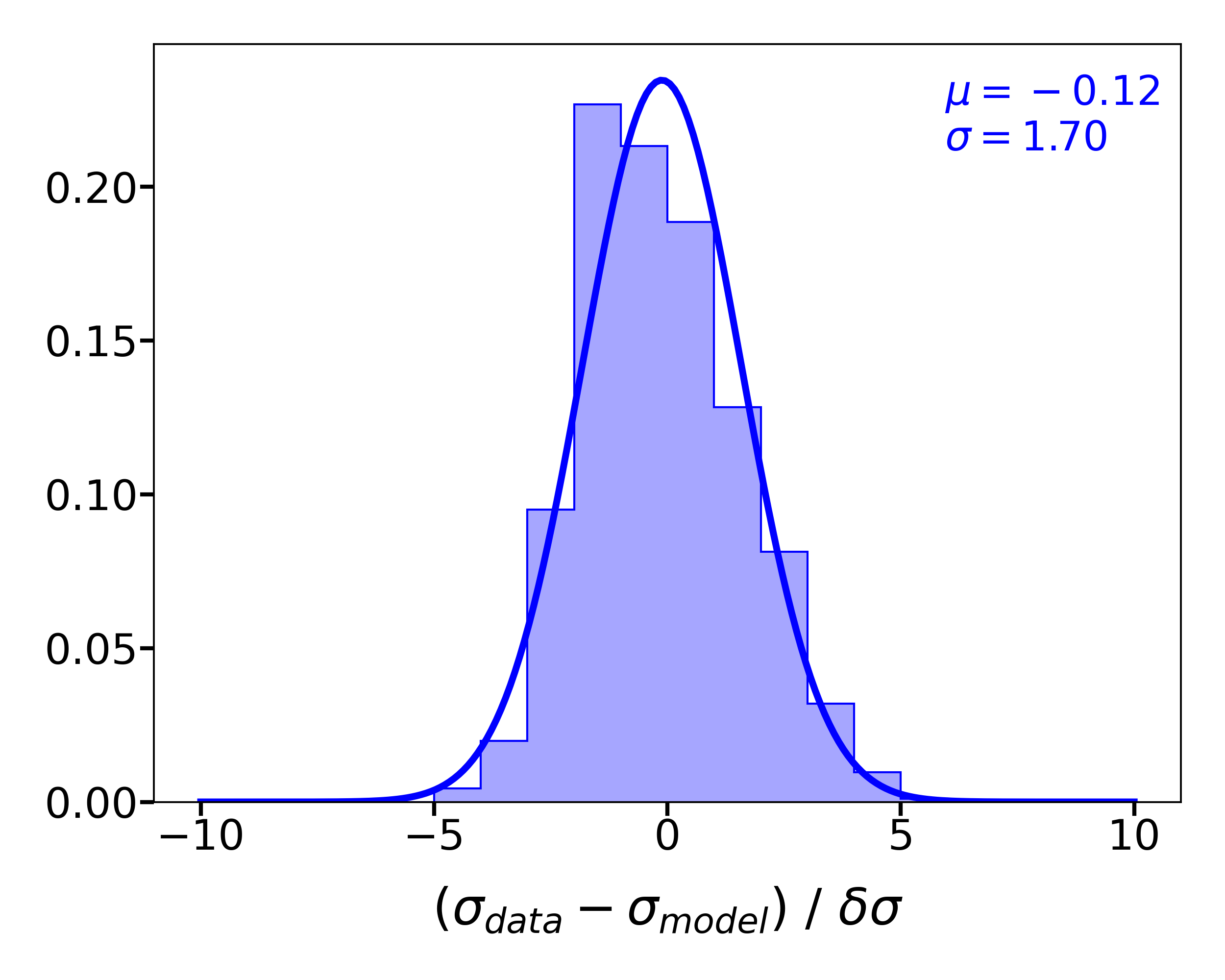}
    
    \caption{Best fit model of the NGC 7683 galaxy. The first, second and third columns refer to the surface brightness, the line of sight velocity and the line of sight velocity dispersion, respectively. From top to bottom we report the observational data, the best fit model, the residual maps and the histograms of the residuals \fr{together with the result of a Gaussian fit}. The residuals are divided by the data errors $\delta v$, $\delta \sigma$ in the case of the velocity and velocity dispersion, and by the data in the case of the surface brightness.}
    \label{fig:best_fit_map}
\end{figure*}

Fig.~\ref{fig:best_fit_map} illustrates the best fit model obtained from our analysis for the logarithm of the surface brightness (left column), the line of sight velocity (middle column), and the line of sight velocity dispersion (right column). The first and the second rows refer respectively to the data and the best fit model, while in the last two rows we show 
the residuals respectively divided by the data errors $\delta v$ (for the velocity), $\delta \sigma$ (for 
the velocity dispersion), and by the data themselves (in the case of the surface brightness). Comparing our model predictions 
with the data, we generally find a quite good agreement between all the three quantities, as can also be noticed by looking at the residuals maps (third row) and histograms (last row), which are relatively small in magnitude and do not exhibit any systematic deviation from zero.

\fr{The histograms are fit with Gaussian functions whose mean ($\mu$) and dispersion ($\sigma$) are reported in fig.~\ref{fig:best_fit_map}. For all three observables $\mu$ is always close to $0$, highlighting the absence of large systematic deviations between model and data. Similarly, $\sigma$ is in general quite small even though, especially for the kinematics, it is larger than unity. This is due to the presence of features and noise in the data which cannot be reproduced by the intrinsically smooth nature of our model.}

\begin{figure*}
    \centering
    \includegraphics[width=\textwidth,height=0.5\textheight]{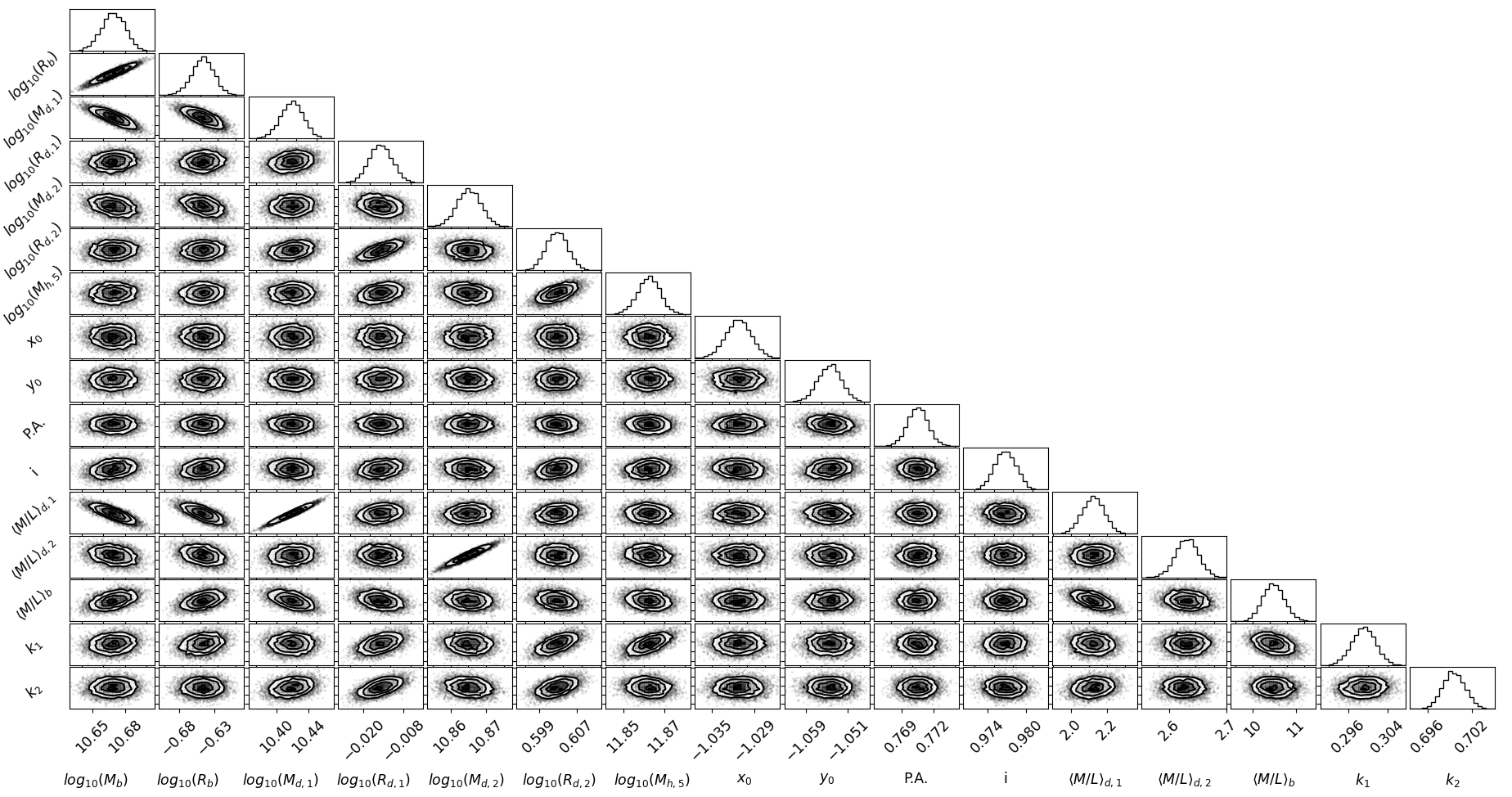}
    \caption{Corner plot of the posterior probability computed from the fit of the NGC 7683 galaxy. The posteriors for each of the parameters are well constrained with an almost Gaussian shape. }
    \label{fig:corner_plot}
\end{figure*}
\begin{figure*}
    \centering
    \includegraphics[width=0.32\textwidth,height=0.19\textheight]{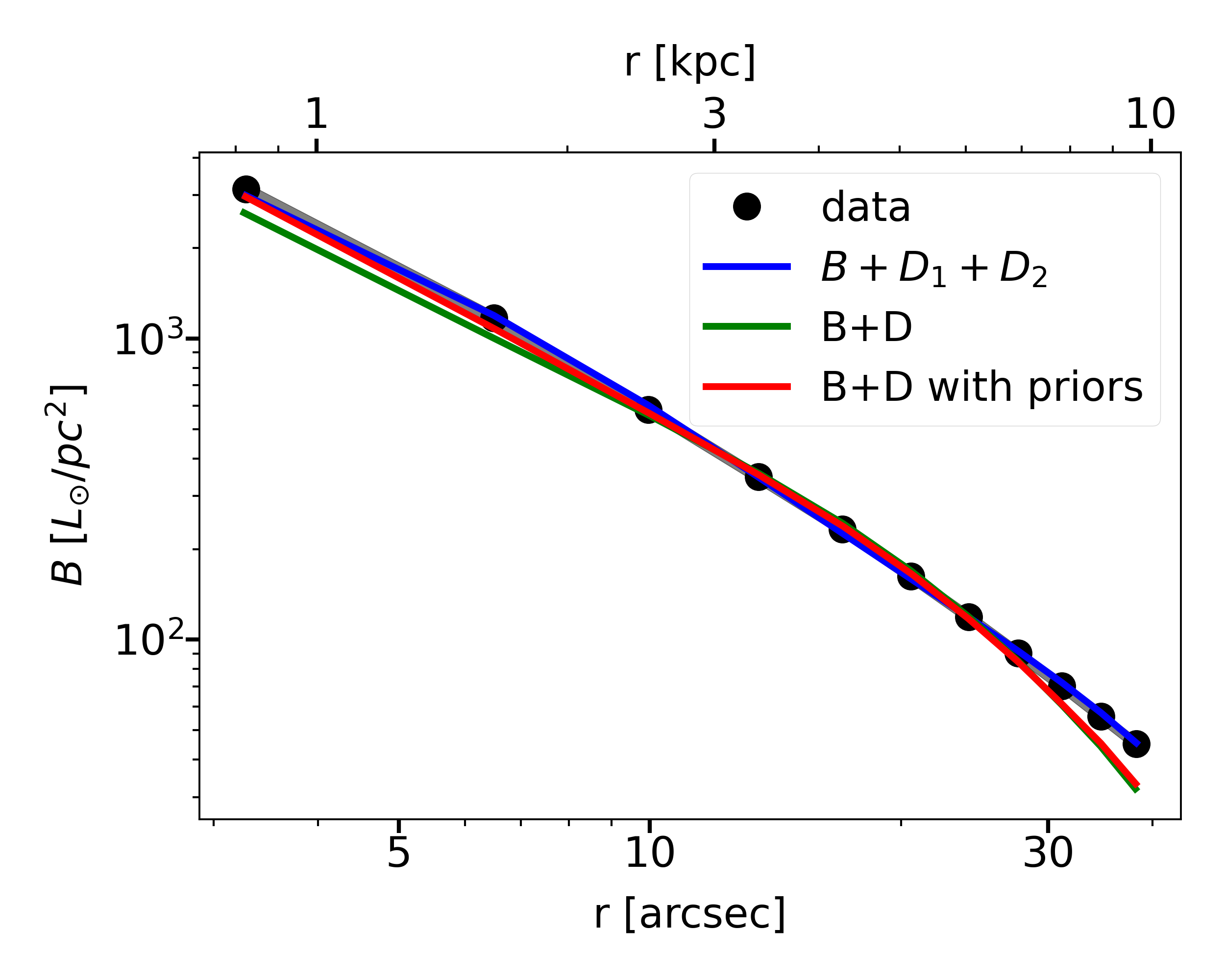}
    \includegraphics[width=0.32\textwidth,height=0.19\textheight]{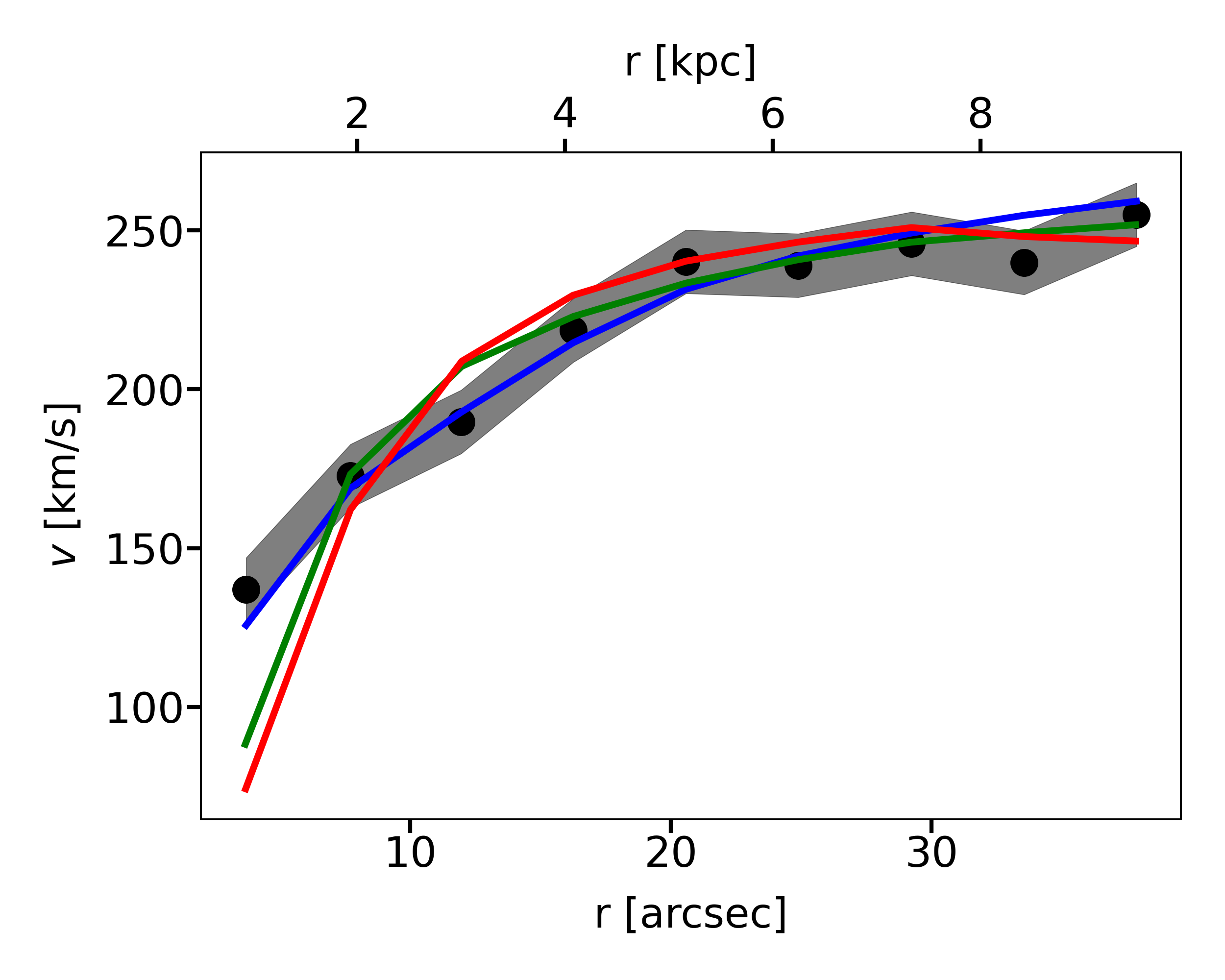}
    \includegraphics[width=0.32\textwidth,height=0.19\textheight]{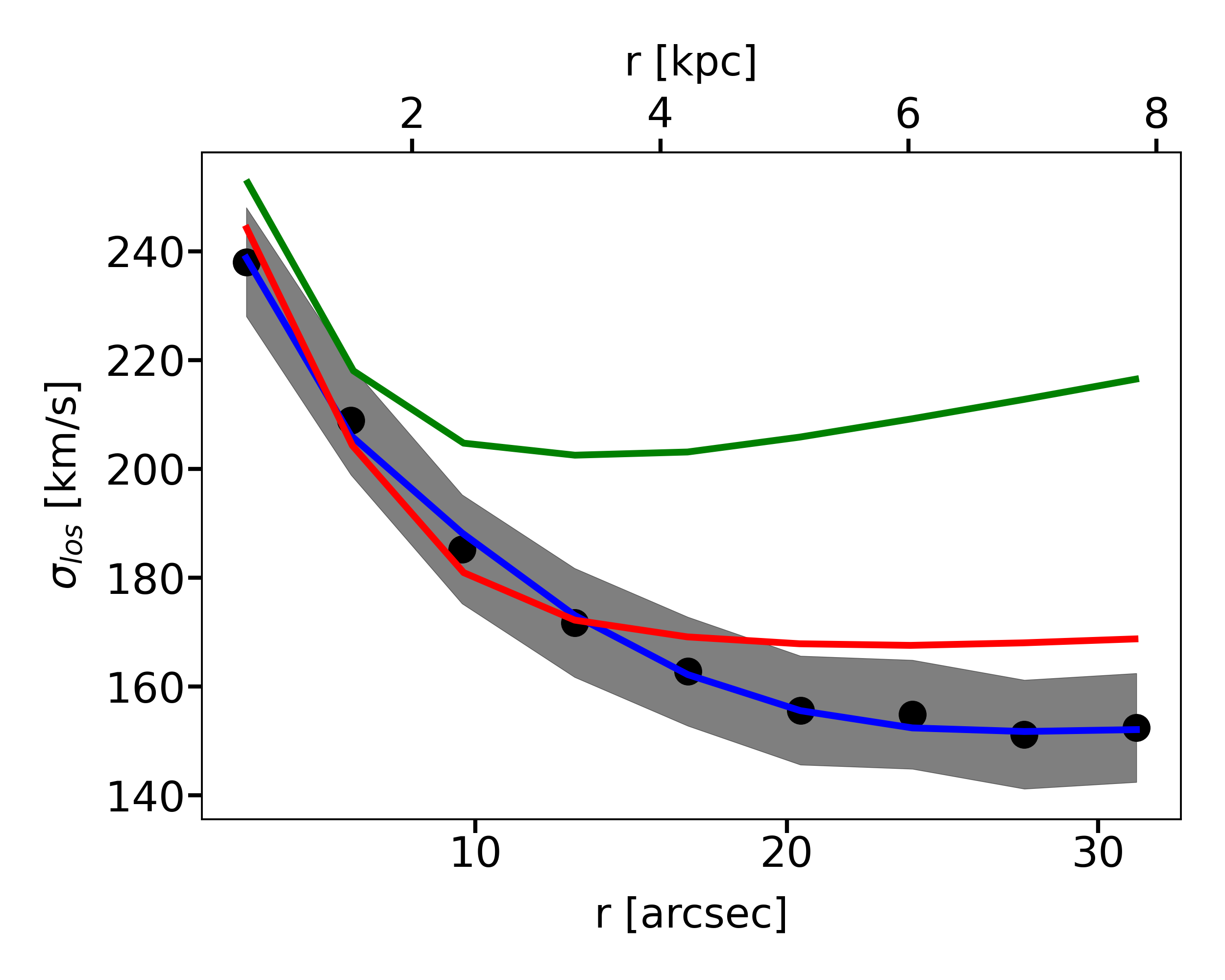}
    \caption{Average radial profiles of the logarithm of the surface brightness (left-hand panel), of the velocity (middle panel) and of the line of sight velocity dispersion (right-hand panel). The data are reported as black dots, with the errors (a 5\% uncertainty for the brightness, $10 \  \mathrm{km/s}$ for the velocity, and $15 \ \mathrm{km/s}$ for the velocity dispersion) are shown as a grey shaded area. The blue lines refer to the the best fit model in the case of three visible components (bulge + inner disc + outer disc), the green lines to a model composed by a bulge and a disc, and the red lines to a bulge plus a fast-rotating disc (priors on $k>0.7$, on $(M/L)_{d}>2.5$ and on $\log_{10}{M_b}<11.1$).}
    \label{fig:radial_profile}
\end{figure*}

The corner plot in Fig.~\ref{fig:corner_plot} describes the behaviour of the posterior probability for all the estimated parameters. More specifically, each column refers to a specific parameter of the model and the top panel represents its one dimensional (i.e. marginalized over all the other parameters) probability distribution. The remaining panels illustrate the joint posterior probability between different parameters and are used to detect possible correlations among them.

The one dimensional histograms reveal a clear peak with an almost Gaussian shape characterised by small widths suggesting optimal convergence for all the reported parameters. As a reference value we report in Tab.~\ref{tab:best_fit_parameters} the medians of the distributions, and the 84th (16th) percentile as upper (lower) uncertainties. Looking carefully at the errors we can see that most of them are often smaller than a few percents; this is a quite common situation when using nested sampling algorithms,\footnote{see for example \citet{Pancoast_2014} which includes a "temperature term" in the likelihood in order to increase the errors on the estimated parameters.} and, in our case, it is probably related to the simplified nature of the model. We stress the fact that the errors reported in Tab.~\ref{tab:best_fit_parameters} rely on the assumption that our model is catching all the relevant physics although, in principle, 
the best fit parameters and their errors should be estimated by marginalising over all the possible dynamical models. Future works based on more complete modelling may result in more reliable uncertainties on the parameters.


The two dimensional histograms reported in the corner plot of fig. \ref{fig:corner_plot} represent the posterior probability of two parameters after marginalising over all the remaining ones. Analysing them in details we can identify some interesting trends. For instance the posterior probabilities of $x_0$, $y_0$ and $\rm P.A.$ are uncorrelated with all the other parameters; these three physical quantities, which define the geometrical location of the system, are indeed the firsts to be constrained during the fit and their estimated value is fairly independent on the other parameters. This behaviour is not unexpected since, for example, the location of the center $(x_0$,$y_0)$ is consistent with the brightest pixel of the galaxy whose position is independent on the "intrinsic" parameters of the model. Other parameters, on the contrary, show very strong correlations as in the case of the $M/L$ ratios and the respective mass. This positive correlation is consistent with the fact that an increase in both these two quantities will result in the same luminosity. 

The most massive component is the outer disc, which is the only one exhibiting a fast rotation ($k\sim0.7$), the k parameter starts from a dispersion dominated component (the bulge) where k is essentially zero and increases gradually up to nearly circular motion. The bulge is a fundamental component in order to account for the large dispersion in the center, but note that its scale radius is of the order of the PSF width, suggesting that it could probably be interpreted, together with the inner disc, as a single structure describing a rotating (pseudo)-bulge.

\begin{table}
	\centering
	\caption{Evidences of the three models reported in Fig.~\ref{fig:radial_profile}.}
	\label{tab:evidences}
	\begin{tabular}{lr} 
		\hline
		Name & $\log{Z}$ \\
		\hline
		$\mathrm{B+D_1+D_2}$ & -24331\\
		$\mathrm{B+D}$ & -31999\\
		$\mathrm{B+D}$ with priors & -36706\\
		\hline
	\end{tabular}
\end{table}

The need for rotational motion in the center of NGC 7683 can also be deduced from a statistical analysis based on the evidence comparison of different dynamical models. Fig.~\ref{fig:radial_profile} shows the average radial profiles of the surface brightness, the velocity and the line of sight velocity dispersion for different dynamical models. The black dots correspond to the data and the grey shaded area to the associated errors.\footnote{Here we assume a 5\% error in the brightness, $\mathrm{10 \ km/s}$ for the velocity and $\mathrm{15 \ km/s}$ for the velocity dispersion.} The blue line is the best fit model assuming a bulge, an inner disc and an outer disc ("$\mathrm{B+D_1+D_2}$"), the green line refers to a model with a central bulge and a single disc component ("B+D"), and the red line to a model with a bulge and a rotation-dominated disc ("B+D" with priors\footnote{We use priors on $k>0.7$, on $(M/L)_{d}>2.5$ and on $\log_{10}{M_b}<11.1$}). All the three models assume the same dark matter component. 

From Fig.~\ref{fig:radial_profile} it is clear that the "$\mathrm{B+D_1+D_2}$" model is a much better fit to the data. Nonetheless, this does not necessarily imply a deeper understanding of the underlying physics. More precisely, the "$\mathrm{B+D_1+D_2}$" model has 4 additional parameters with respect to the other models and, when models with a different number of parameters are compared, the residuals are not always a sufficient statistical diagnostic to prefer one model with respect to others. In our case, however, the algorithm also computes the evidences of the three models (see Tab.~\ref{tab:evidences}) clearly indicating that the one with the (by far) highest $Z$ (i.e. "$\mathrm{B+D_1+D_2}$") is statistically preferred, and confirming that the addition of the second disc component is well motivated \footnote{\fr{The \textit{odds ratio} computed as the fraction of the evidences of the "$\mathrm{B+D_1+D_2}$" model in respect to the others is $\gg$ 1 meaning that, from a statistical point of view, the "$\mathrm{B+D_1+D_2}$" is the most favoured one.}}.

Comparing the evidences of the "$B+D$" and the "$B+D$" with prior models we can see that the first one is significantly better, even though the details of the velocity dispersion profile are not reproduced at all (as can be clearly seen in Fig.~\ref{fig:radial_profile}). 
The reason is that the term in the likelihood related to the velocity dispersion has a relatively small weight, especially in the outer regions where the error is significantly higher, hence the nested sampling prefers a solution which fits slightly better the brightness and the velocity in the central regions at the expense of the velocity dispersion.

\begin{figure*}
    \centering
    \includegraphics[scale=0.093]{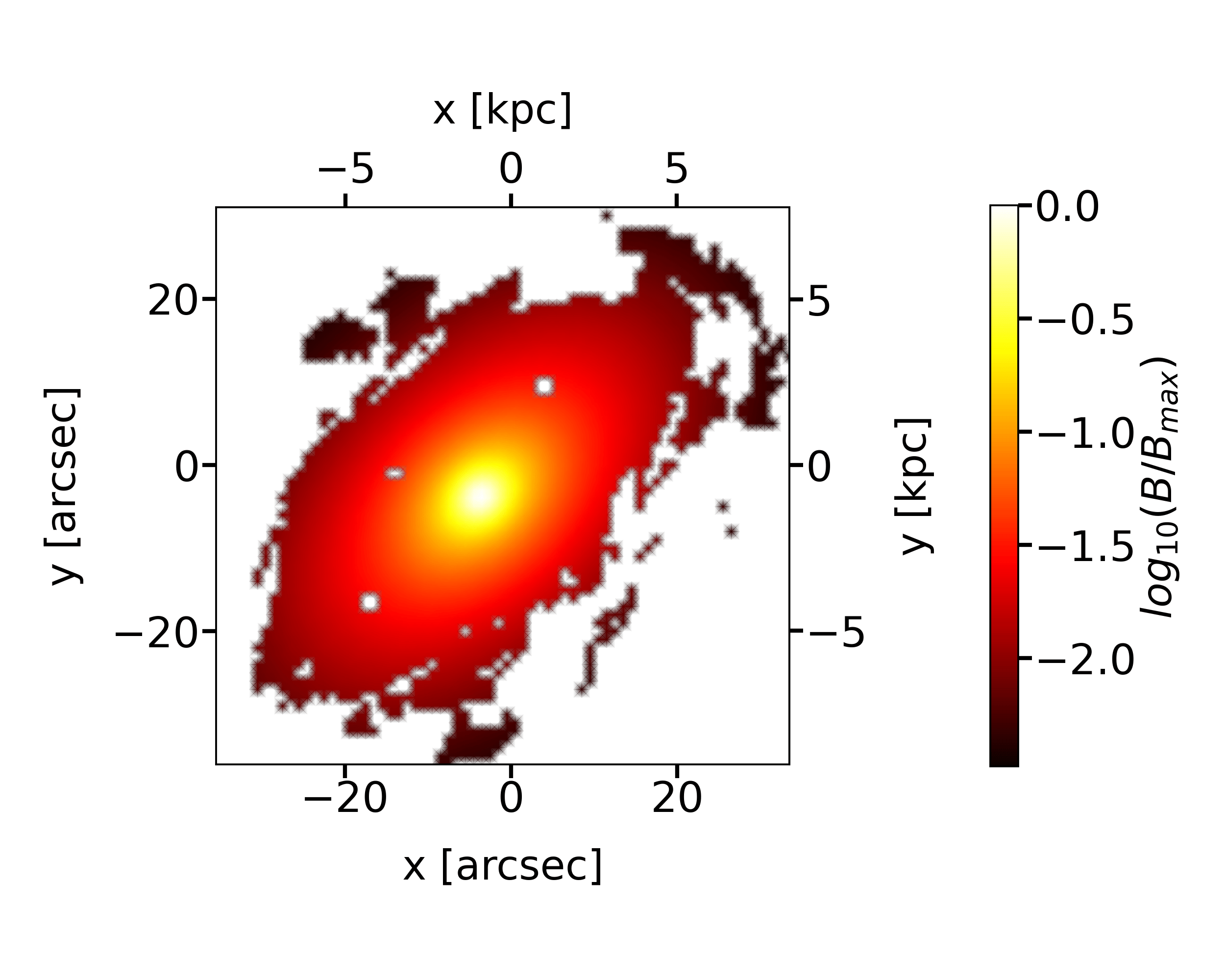}
    \includegraphics[scale=0.093]{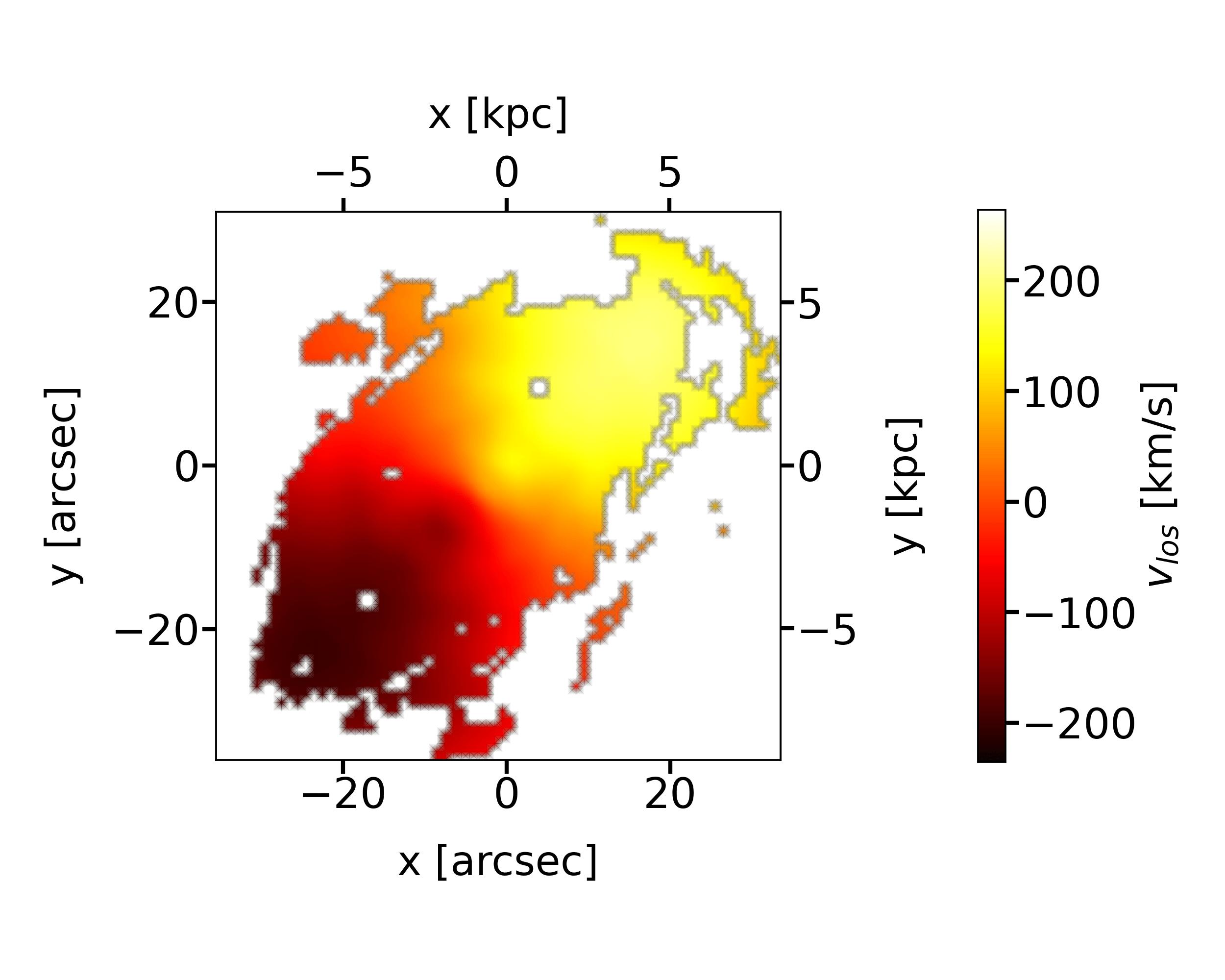}
    \includegraphics[scale=0.093]{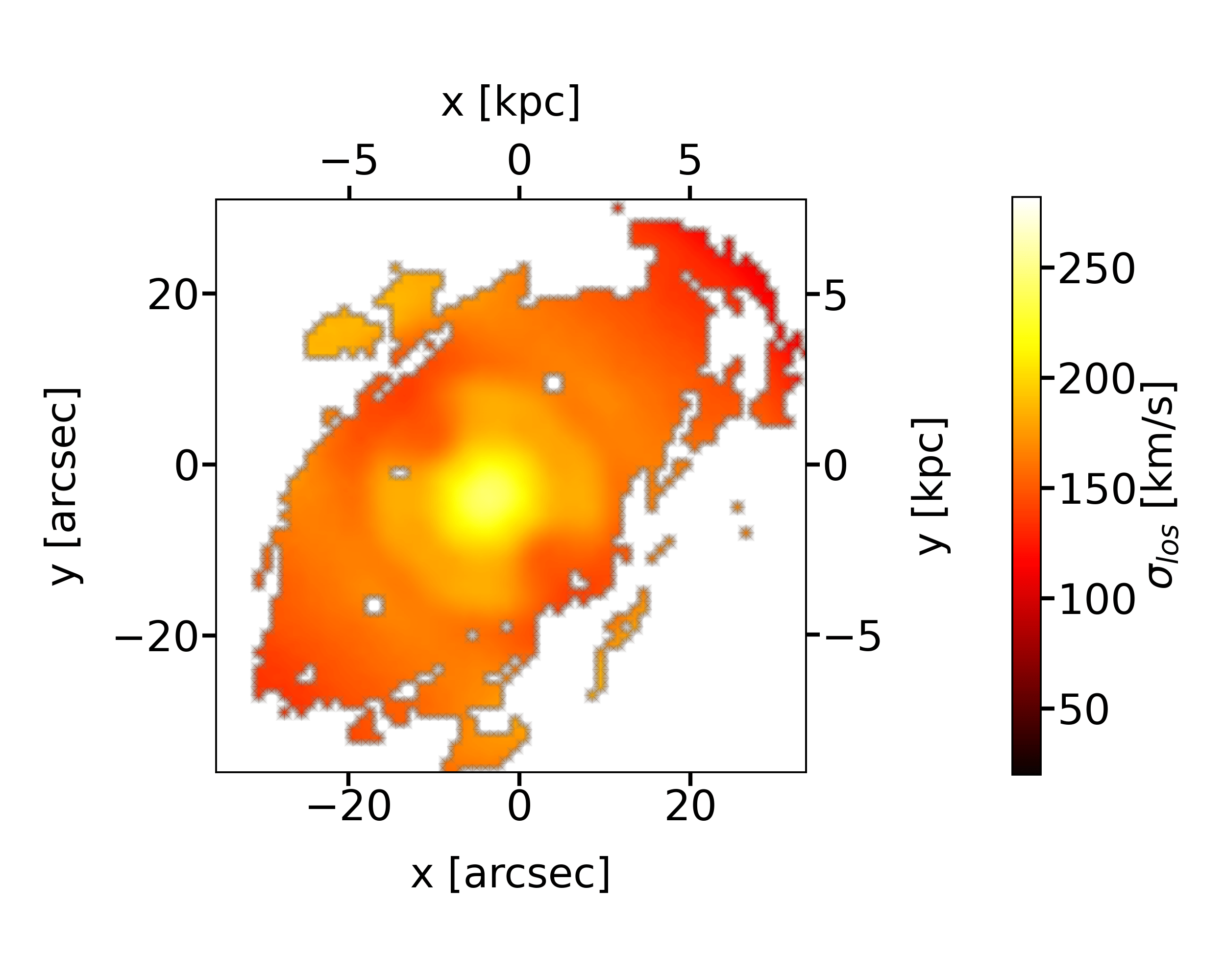}
    \includegraphics[scale=0.093]{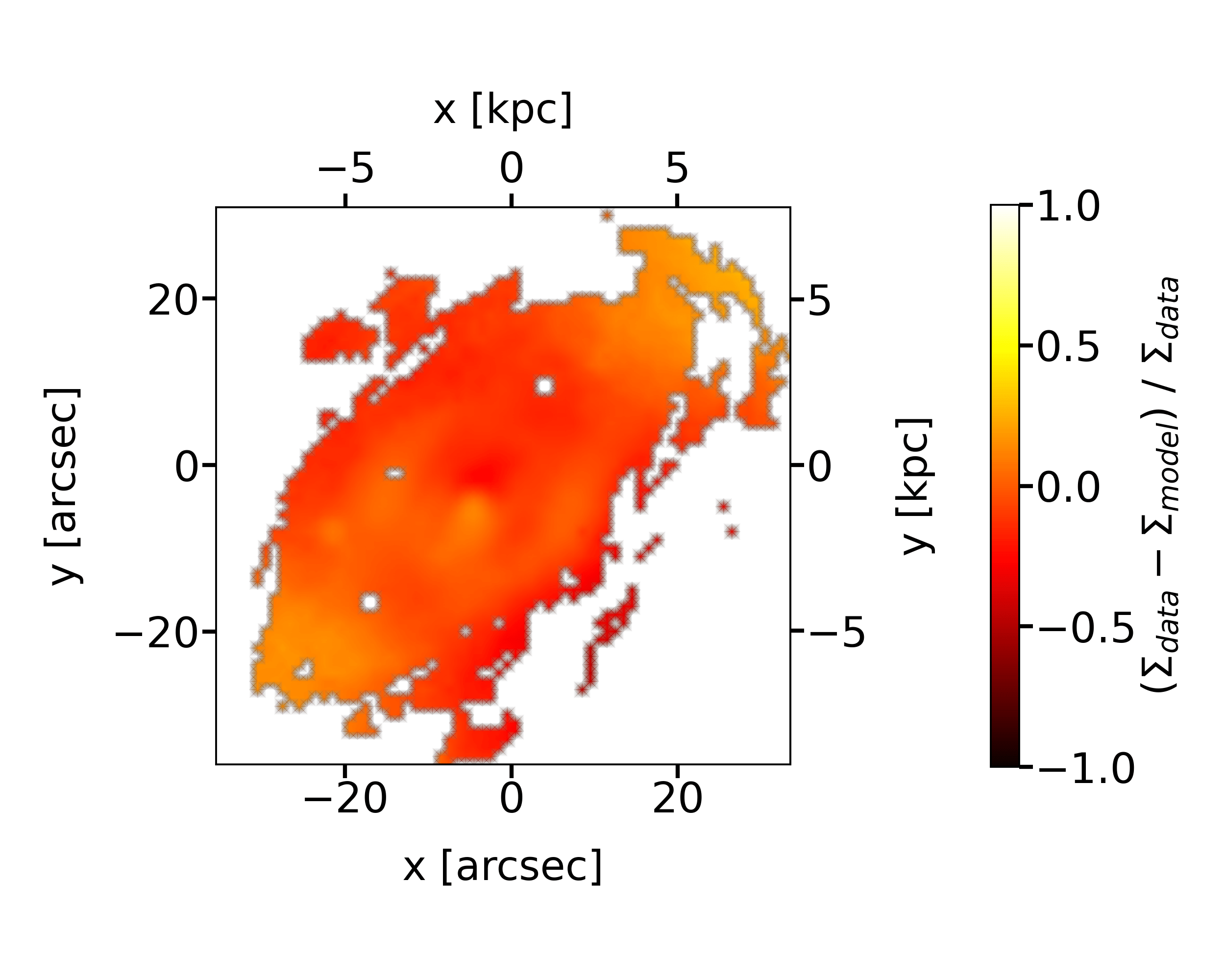}
    \includegraphics[scale=0.093]{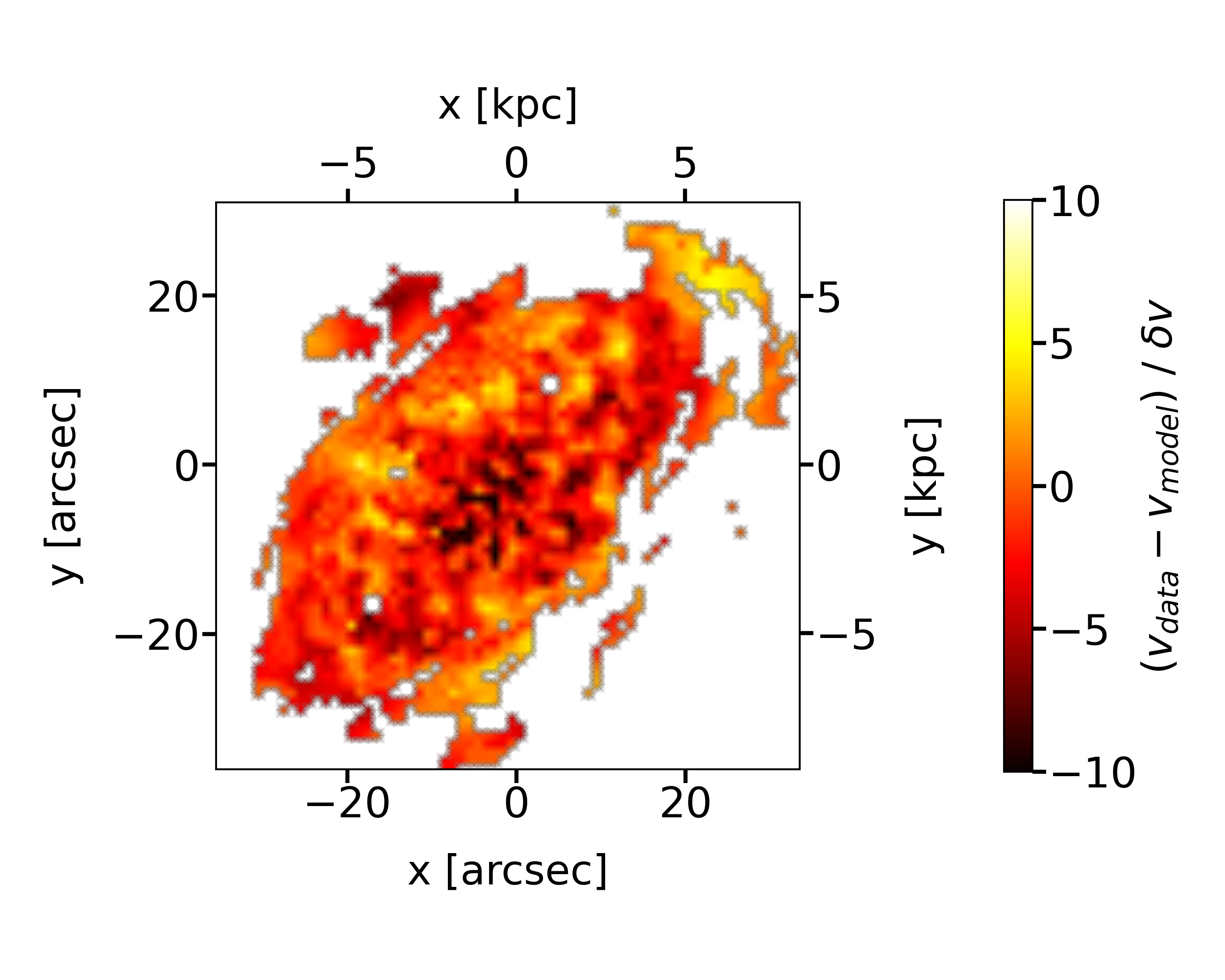}
    \includegraphics[scale=0.093]{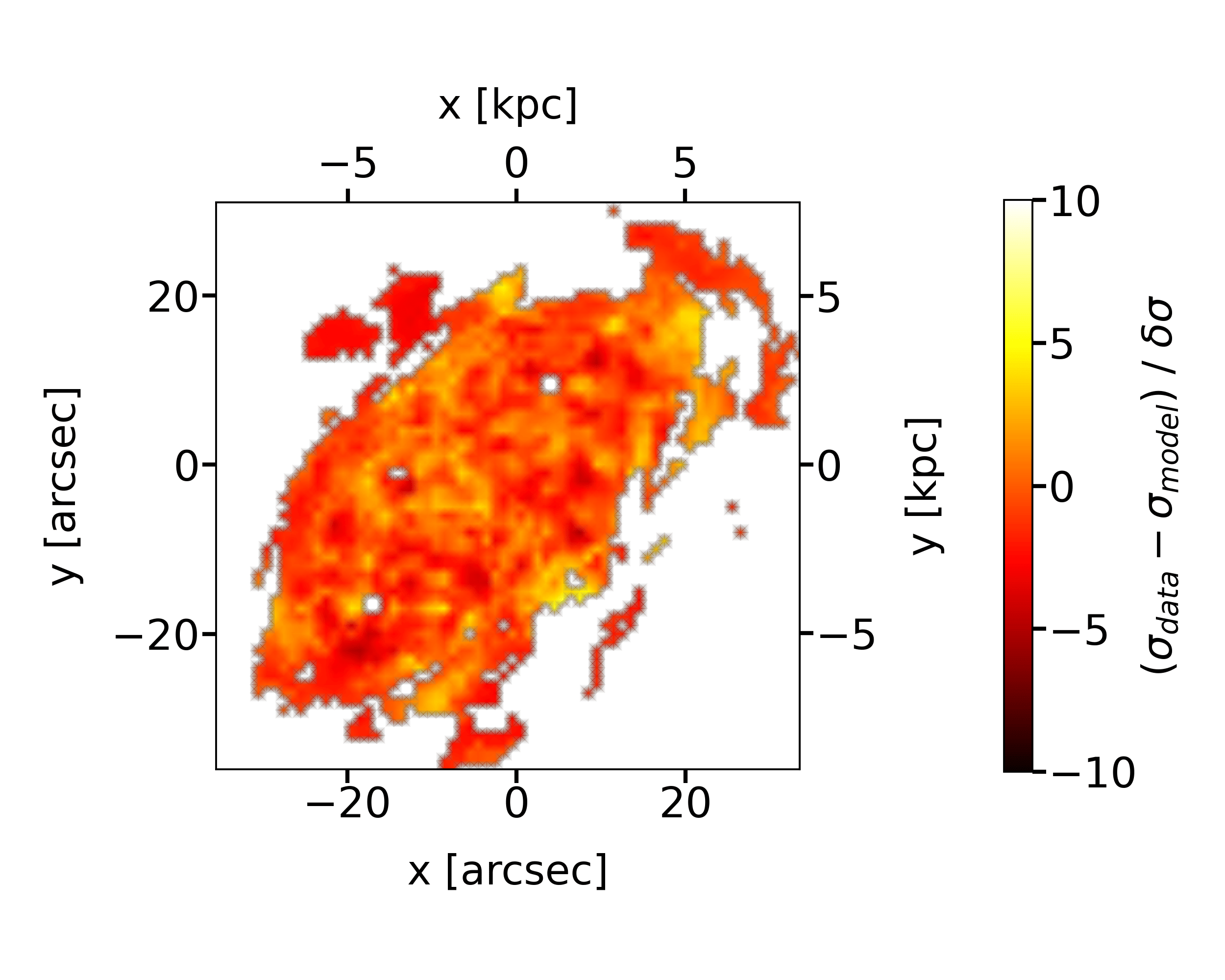}\\
    \hspace{-0.075\textwidth}
    \includegraphics[width=0.23\textwidth,height=0.18\textheight]{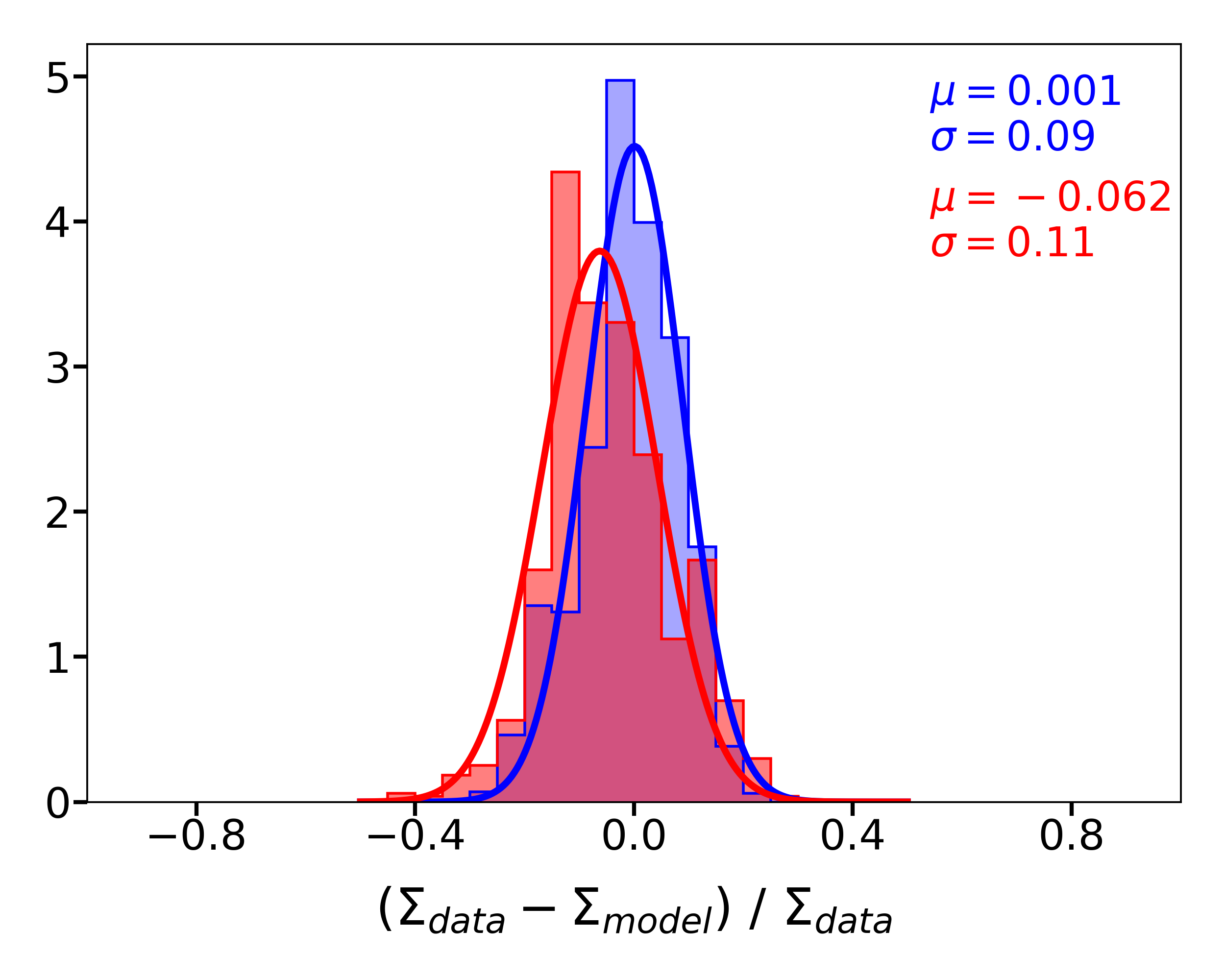}\hspace{0.09\textwidth}
    \includegraphics[width=0.23\textwidth,height=0.18\textheight]{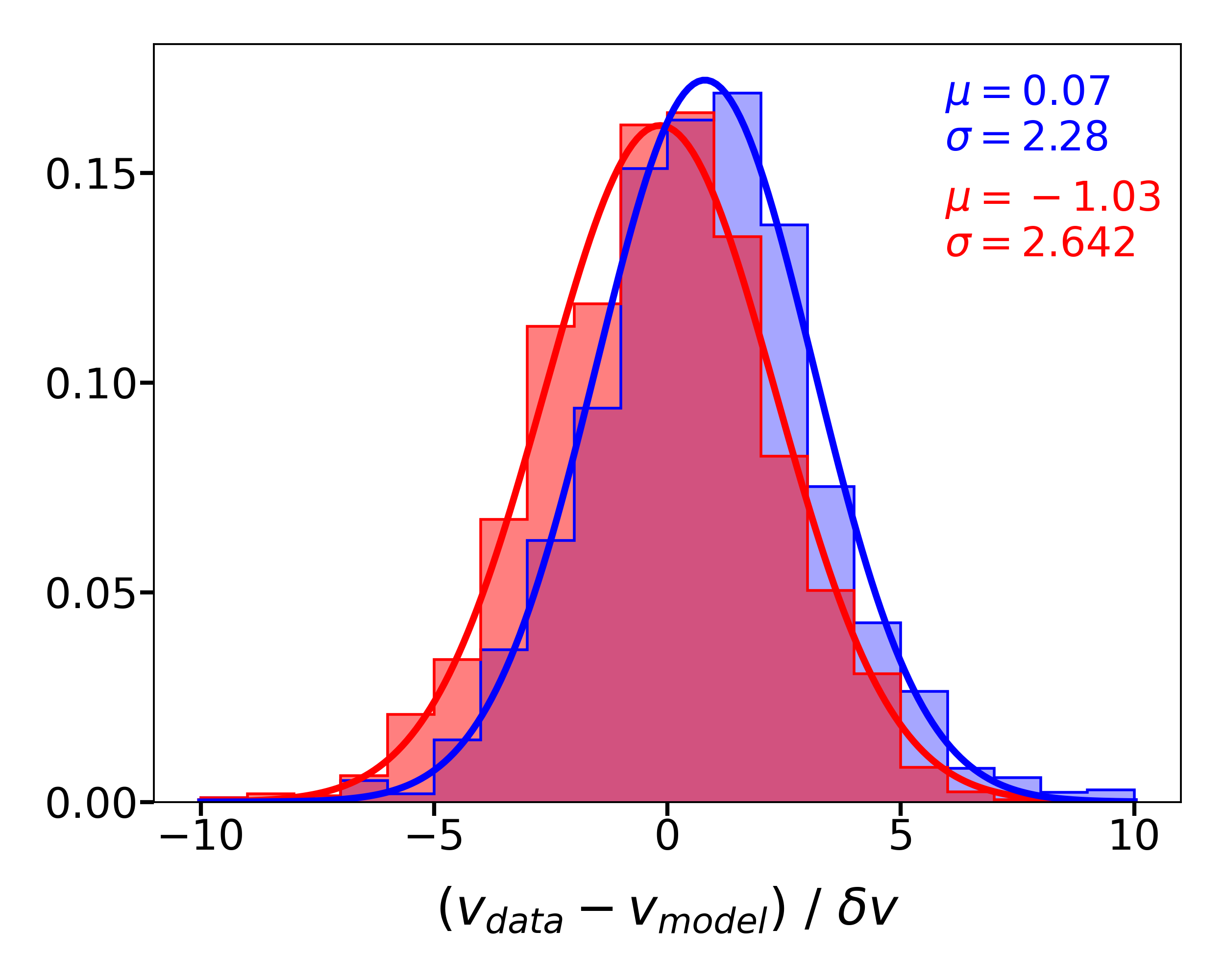}\hspace{0.105\textwidth}
    \includegraphics[width=0.23\textwidth,height=0.18\textheight]{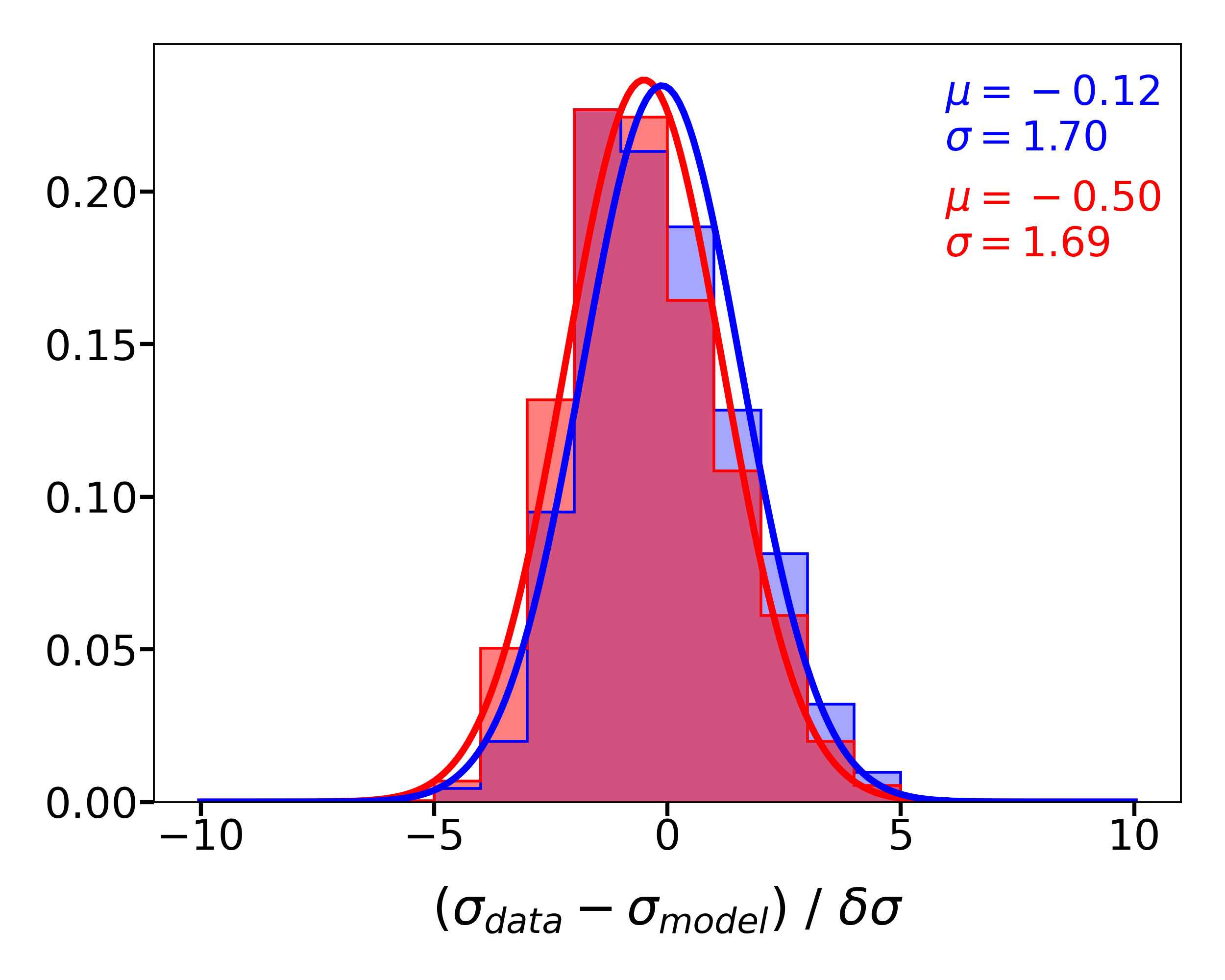}
    \caption{Best fit model obtained with \textsc{dynamite} for the NGC 7683 galaxy. The first column refers to the surface brightness, the second column to the line of sight velocity, and the last one to the line of sight velocity dispersion. From top to bottom, we report the best fit model, the residual maps, and the histograms of the residuals \fr{together with Gaussian fits of the 2 distributions. The blue histograms refer to the residuals of our model while the red ones come from the \textsc{dynamite} model) .}}
    \label{fig:best_fit_map_dynamite}
\end{figure*}

We also want to stress that the "B+D" models are clearly underestimating the amount of rotation in the center of the NGC 7683 galaxy, due to the presence of the isotropic bulge (see the middle panel in Fig.~\ref{fig:radial_profile}). Our statistical analysis suggests that the inner disc is crucial in order to account for significant rotation in the central regions of the galaxy, strengthening the interpretation that the bulge + inner disc system can be considered a reasonable approximation for a rotating central structure such as a pseudo-bulge or the remnant of a former bar \citep[see e.g.][]{Kormendy_2004}. Such structures are currently not implemented in our models, but are a natural extension that we plan to include in a future follow-up.

\subsection{Comparison with DYNAMITE}
\label{sec:comparison_with_dynamite}
In this section we compare the results described in the previous paragraph with a similar analysis performed with \textsc{dynamite}
\citep{DYNAMITE}, a publicly available tool based on the Schwarzschild orbit superposition method \citep{Schwarzschild1979}, useful for investigating the internal properties of galaxies.

The Schwarzschild method can be schematically divided in three main steps:
\begin{figure*}
    \centering
    \includegraphics[width=0.32\textwidth,height=0.19\textheight]{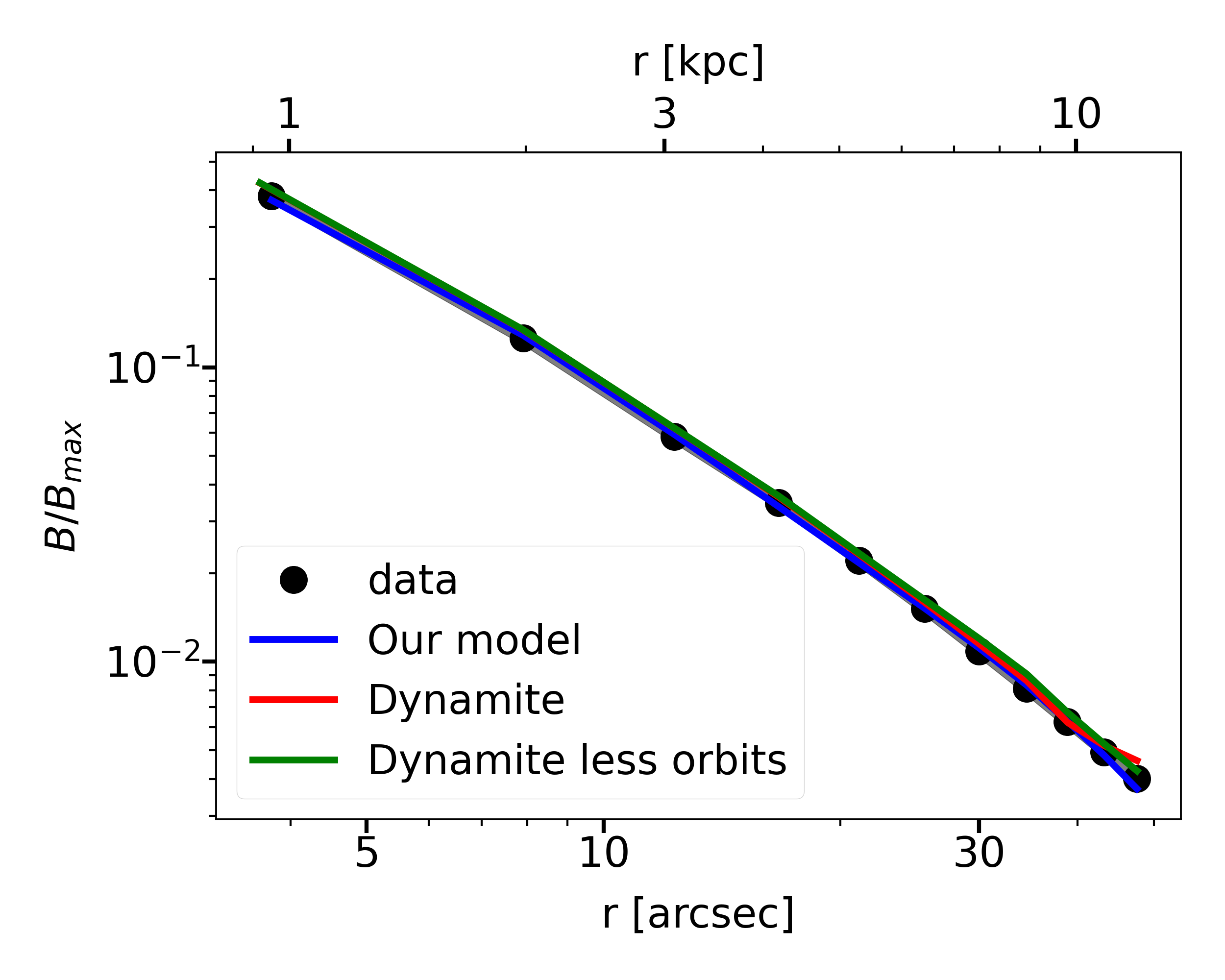}
    \includegraphics[width=0.32\textwidth,height=0.19\textheight]{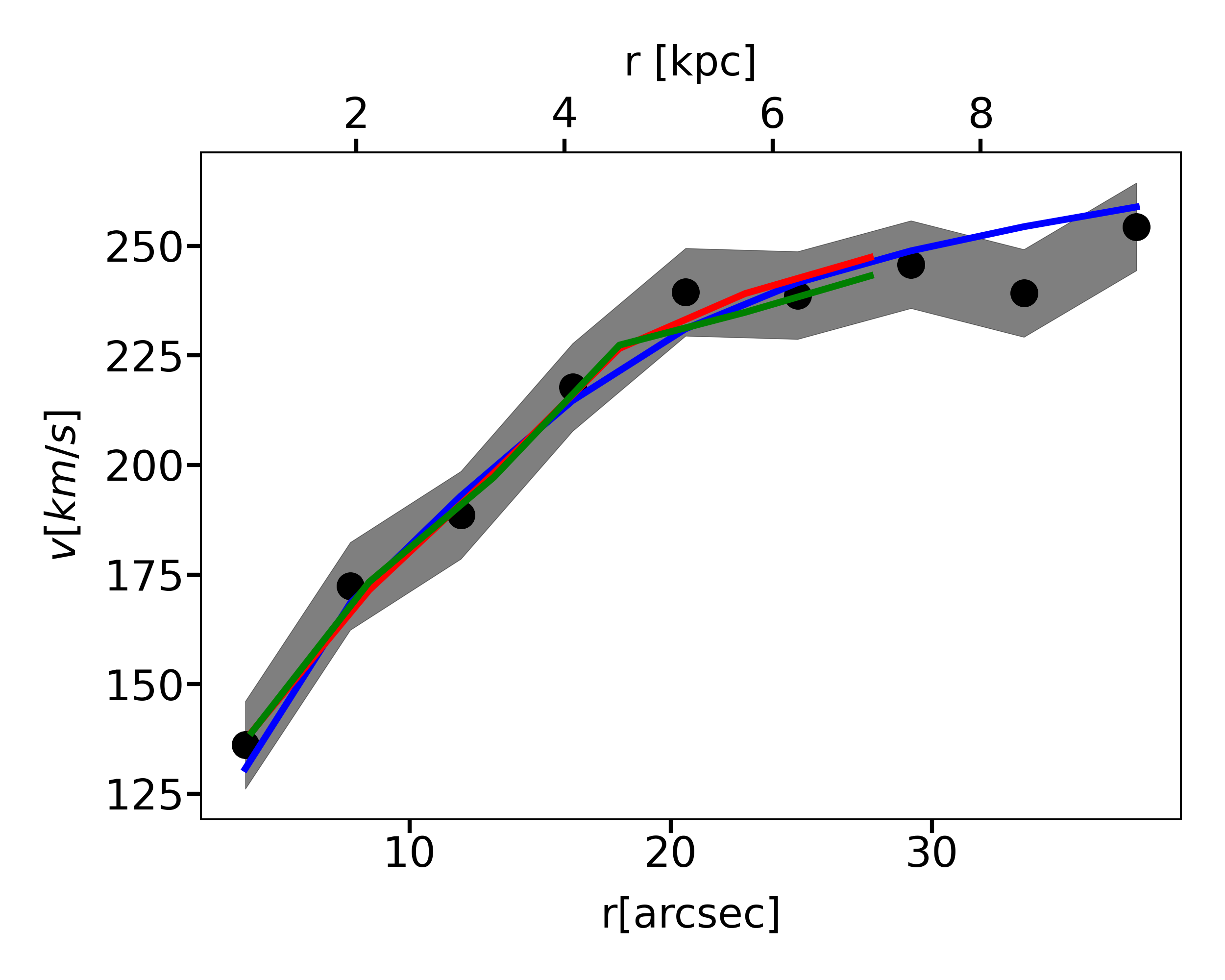}
    \includegraphics[width=0.32\textwidth,height=0.19\textheight]{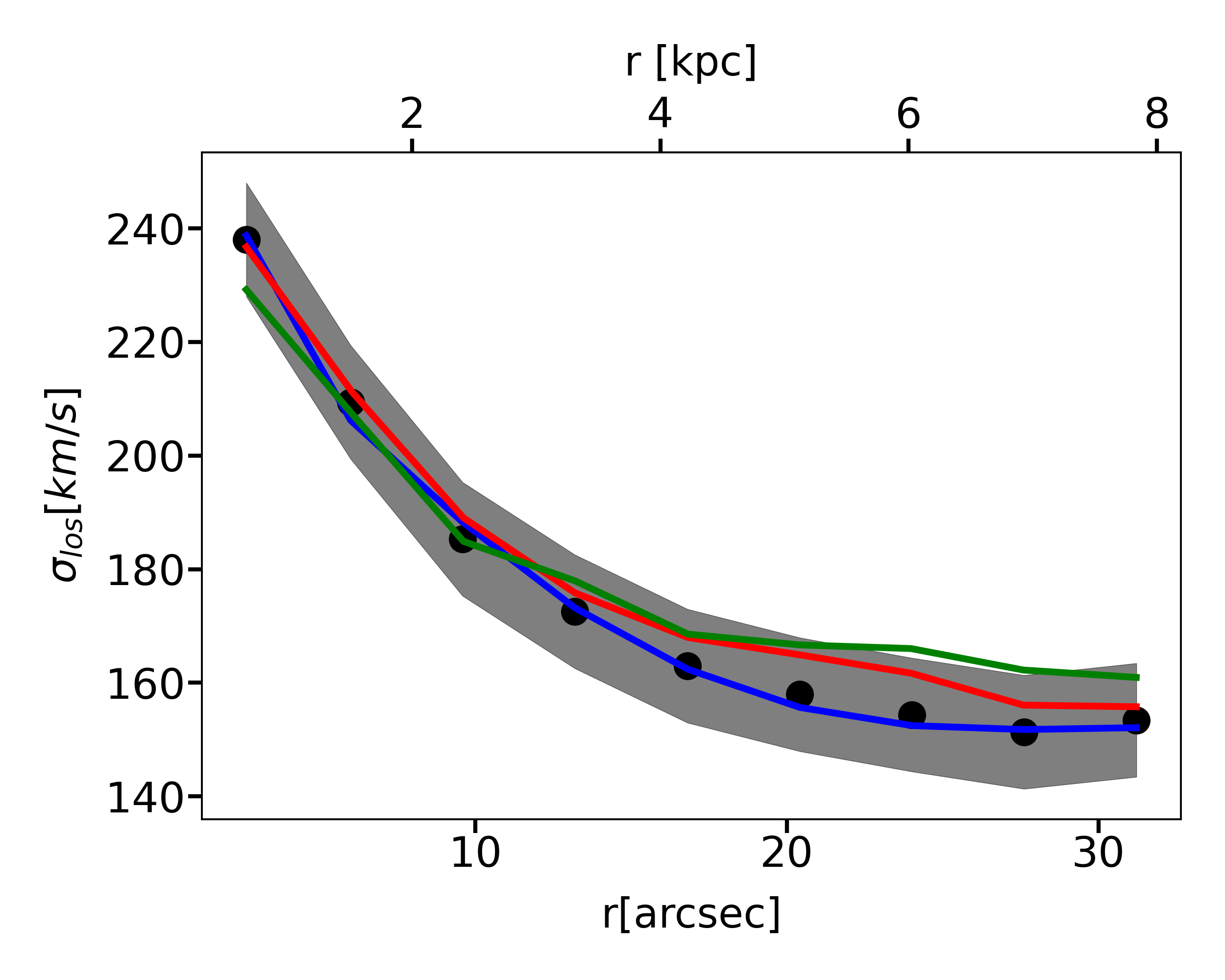}
    \caption{Same as Fig.~\ref{fig:radial_profile}, but comparing our best model (shown as blue lines) with \textsc{dynamite}. The red and green lines correspond to the best fit model obtained with \textsc{dynamite} using a different number of orbits, i.e. 285 000 orbits and 95 000 orbits respectively.}
    \label{fig:radial_profile_dynamite}
\end{figure*}
\begin{itemize}
    \item The first step consists in determining the total potential, usually a superposition of dark components (halo and/or black hole) and at least one visible component, assumed to be triaxial. The visible components are typically described by superpositions of triaxial Gaussian ellipsoids (MGE, \citealt{Cappellari_2002}), whose 2-D ellipsoidal Gaussian projections depend on four parameters: their total luminosity ($L_j$), the axial ratio ($q_j'$), the dispersion along the major axis ($\sigma_j$),\footnote{The dispersion about the minor axis is then univocally determined by the $q_j'$ parameter.} and the position angle ($\psi_j$), where $j$ refer to the specific Gaussian ellipsoid. These parameters are fitted to the observed surface brightness maps \footnote{Note that for the MGE we used a larger field of view in respect to the analysis in the previous paragraph with the full SDSS spatial resolution.} using a large enough number of Gaussian components. Sometimes it can be useful to give some physically motivated constraints to the parameters, for instance in our case we used a constant position angle and fixed $q_j'>q'_{min}=0.6$. As dark component (contributing only to the potential) we selected among the 4 possible choices a Navarro-Frenk-White dark matter halo with a fixed concentration parameter 
    $c=8$, in order to reduce the dimension of the parameter space. Due to the limited spatial resolution of our data, we did not include any central massive black hole in the model. \textsc{dynamite} requires as input a `dynamical' mass-to-light ratio ($(M/L)_{\rm dyn}$) which is a multiplication factor of the overall potential. This prevents the re-computation of all the orbits (see next step) when only $(M/L)_{\rm dyn}$ varies. We decided instead to fix $(M/L)_{dyn}=1$ and to define a stellar mass to light ratio $(M/L)_{\star}$ which is directly applied on the MGE. 
    \item In the second step, a huge amount of orbits is integrated for a timescale much larger then the orbital period. In our case, we choosed to integrate approximately 285 000 orbits, each
    for 200 orbital periods. After that, each orbit is properly weighted in order to match the measured moments of the line of sight velocity distribution (LOSVD).\footnote{The LOSVD is usually expanded in terms of Gauss-Hermite polynomials \citep{Cappellari_2016} up to the fourth moment ($h_4$). However, this parametric expansion must be treated carefully since in some cases it can lead to negative and unphysical LOSVDs.} We manually tuned some of the hyperparameters of the weight solver in order to reach optimal convergence in the fit. 
    \item the last step consists in running different dynamical models, in order to gauge all the free parameters which, in our case, are: the logarithm of the halo mass fraction $\log_{10}{f}$, the stellar mass to light ratio $(M/L)_{\star}$ and the intrinsic flattening $q$. The intrinsic flattening is defined as the ratio between the major axis and the minor axis of the triaxial Gaussian ellipsoid (\citealt{Cappellari_2002}). To explore the parameter space, we created a $9\times9\times4$ grid covering respectively the ranges: $(M/L)_{\star} \in [2.85,4.15]$, $q \in[0.3,0.59]$ and $\log_{10}{f} \in [0.9,1.8]$. In order to allow for a fast exploration of all the $9\times9\times4$ parameter combinations, we used a high-performance-computing machine equipped with 4 nodes of 48 cores each, resulting in 192 models evaluated simultaneously. The evaluation of a single model lasts for hours, and the occupied disc memory is about $2.5\rm GB$ (a total of $\sim 800 \rm GB$). Note that in total we explored 324 different models, which is a number significantly smaller then the amount of likelihood evaluations required by any Bayesian technique; for these reason  the parameter estimation performed with the grid search should be treated carefully.
\end{itemize}

Fig.~\ref{fig:best_fit_map_dynamite} shows the best fit model, which has been selected using the minimum kinematic $\chi^2$, with parameters: $q=0.4$, $(M/L)_{\star}=4.1\rm\, M_{\sun}/L_{\sun}$ and $\log_{10}f=1.125$. Inverting Eq.~(9) of \citep{Cappellari_2002}, we can estimate the inclination angle of the galaxy as $i=60\,\deg$, which is in agreement with our estimation of $i=55.69^{+0.06}_{-0.06}\, \deg$ considering that we explored only 4 intrinsic flattening with \textsc{dynamite}. Also our estimate for the total visible mass of the galaxy which is $M=1.46^{+0.02}_{-0.02}\times10^{11}M_{\sun}$, is consistent with the value given by the MGE within $40\arcsec$ of $1.5\times10^{11}M_{\sun}$. \fr{A good agreement is maintained also for the total mass profile, although discrepancies exist when separating the visible and dark components. These differences can be associated with $M/L$ ratios which, differently from our model, we assumed constant in the analysis with \textsc {dynamite}.}

As mentioned above, Fig.~\ref{fig:best_fit_map_dynamite} shows the best results obtained with \textsc{dynamite}. The picture scheme is similar to the one reported in Fig.~\ref{fig:best_fit_map}. Note however that the surface brightness is normalized to its maximum and that the dimension of the maps is down-scaled to the dimension of the velocity dispersion one since, differently from our tool, \textsc{dynamite} cannot read as input observables mapped with a different number of pixels.

The residual maps in the second line of the figure show a general agreement between the model and the data, even though the line of sight velocity seems to be slightly overestimated in the center. Also the pattern of the velocity dispersion in the central regions is quite peculiar, with an initial decrease followed by an increase in the outer regions along the minor axis. 
\fr{Despite some little systematic deviations in the surface brightness and in the velocity, the residuals histograms obtained from \textsc{dynamite} (red) and from our model (blue) show  similar scatters ($\sigma$), suggesting that, overall, our simplified model can recover all the observed properties with good accuracy}.

The average radial profiles are presented in Fig.~\ref{fig:radial_profile_dynamite}. The blue line, the black dots and the grey area are identical to Fig.~\ref{fig:radial_profile}, while the red and the green lines refer to the best fit model obtained with \textsc{dynamite} using respectively 285 000 orbits and 95 000 orbits.  Interestingly, we can notice that the modelling of the velocity dispersion improves when the number of orbits is increased. Since these two models have similar best fit parameters, it is not clear whether further increasing the number of orbits (above a certain threshold) is useful to strengthen the 
constraints on the free parameters, or if it 
only adds more degrees of freedom, resulting in smaller residuals. On the contrary, the statistical gain or loss obtained adding new parameters can be gauged in our nested-sampling-based approach. 

\begin{figure}
    \centering
    \includegraphics[width=0.40\textwidth,height=0.27\textheight,trim=5cm 0cm 5cm 0cm]{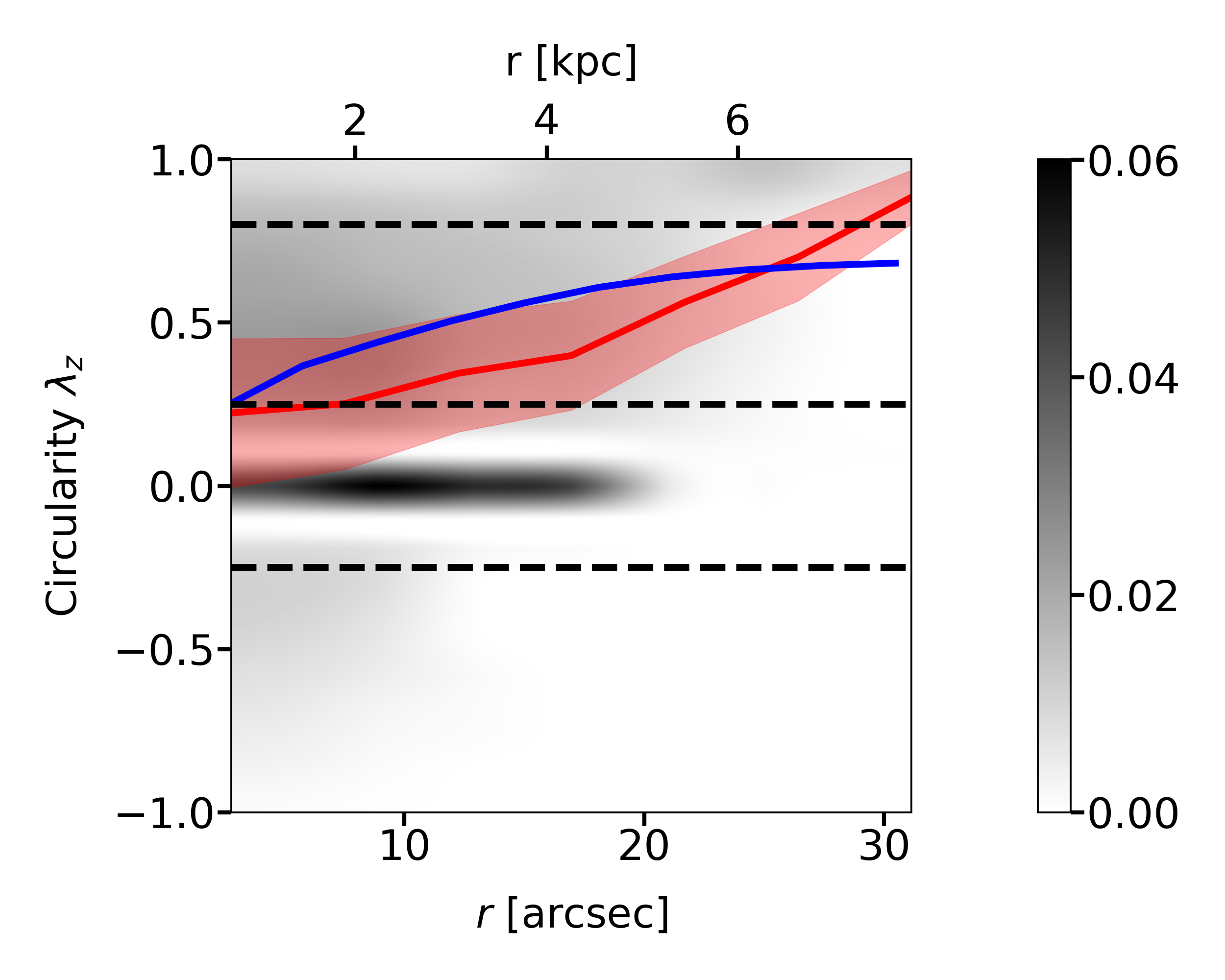}
    \caption{x-axis: distance from the center, y-axis: circularity (i.e. angular momentum along z divided by the angular momentum of the circular orbit). The grey map represents the density of orbits at a specific distance with a given circularity. The red line is the circularity averaged on the orbit density and the shaded area is the associated error computed as the weighted dispersion. The blue line is the circularity predicted from our model which essentially correspond to the average radial profile of the k parameter weighted on the surface brightness. \fr{The horizontal dashed lines split the diagram into different regions depending on the orbit type. From bottom to top: counter rotating orbits, hot dispersion-dominated orbits, warm orbits and cold rotating orbits.}}
    \label{fig:Circularity_dynamite}
\end{figure}

\begin{figure}
    \centering
    \includegraphics[width=0.40\textwidth,height=0.27\textheight,trim=8cm 0cm 2cm 0cm]{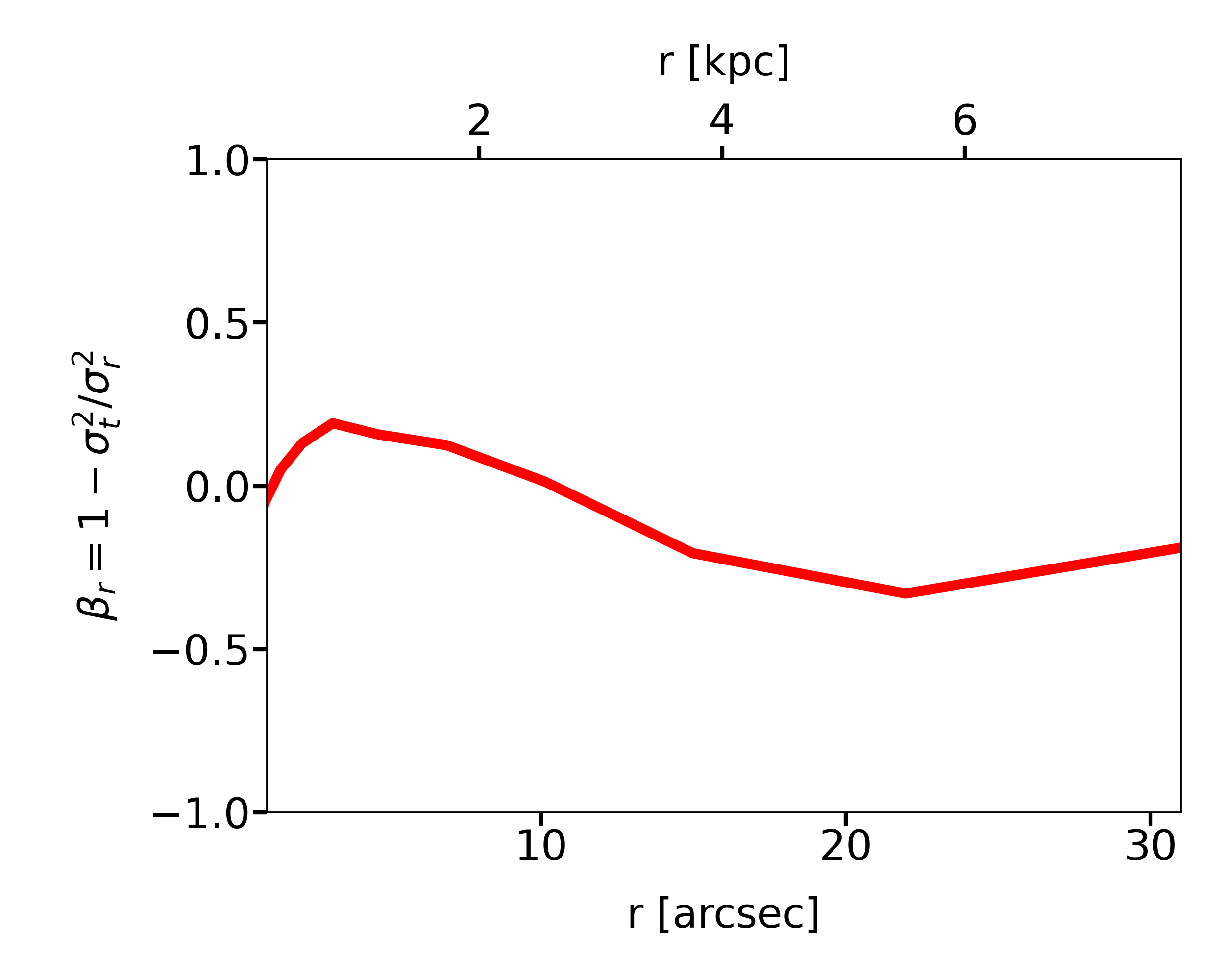}
    \caption{\fr{x-axis: distance from the center,y-axis: anisotropy in spherical coordinates. The anisotropy is computed as $\beta_r = 1-\sigma_t^2/\sigma_r^2$ where $\sigma_t^2 = (\sigma_{\theta}^2+\sigma_{\phi}^2)/2$. $\beta_r = 0$ means isotropy, while $\beta_r>0$ ($<0$) means radially (tangentially) biased orbits.}}
    \label{fig:Beta_dynamite}
\end{figure}

Fig. \ref{fig:Circularity_dynamite} illustrates in grey-scale the density of orbits in the distance vs circularity\footnote{The circularity is defined as the angular momentum along the $z$-axis divided by the angular momentum of a circular orbit with the same energy. In our model, it coincides with the average of the $k$ parameters weighted on their surface brightness} plane, with the red line corresponding to the average circularity weighted by the orbit density from \textsc{dynamite}, the red area to the standard deviation error, and the blue line to the azimuthally-averaged circularity predicted by our model. The figure reveals the presence of 2 clumps with zero circularity superimposed to many rotating orbits. The first groups of orbit with $\lambda_z=0$ is concentrated in the center even though the average profile (red line) does not reach zero, meaning that a non negligible amount of rotation is required also in the central regions, consistently with our statistical analysis described in Section~\ref{sec:our_analysis}. The second bundle of orbits with null circularity is between $15\arcsec$ and $20\arcsec$ and it is probably responsible for the excess in the velocity dispersion seen in the right picture of Fig.~\ref{fig:radial_profile_dynamite}. Note also that the average circularity computed from our model is broadly consistent with the prediction of \textsc{dynamite} even though the two increasing trends have opposite concavity, note also that the most significant difference between the 2 profiles is located in correspondence of the second bundle of orbit with zero circularity.    

\fr{To further check our assumption of isotropy, we show in fig.~\ref{fig:Beta_dynamite} the anisotropy parameter $\beta_r = 1-\sigma_t^2/\sigma_r^2$ as a function of the distance from the center. Isotropy is defined by the condition $\beta_r\simeq 0$, while for $\beta_r>0$ ($<0$) the galaxy is said to be radially (tangentially) biased. The resulting $-0.3 \lesssim \sigma_t/\sigma_r \lesssim  0.15$ supports the assumption of general isotropy. Note, however, that orbits in the center ($r\lesssim 10''$) tend to be radially biased, reaching $\beta_r=0$ only at small radii. Interestingly, the transition to tangentially biased orbits happens at $r\simeq 12''$, where the rotational support and the warm orbits start to dominate.}

\section{Summary and Conclusions}

In this work we described a numerical methodology aimed at the estimation of the global properties of disc galaxies. Our methodology is novel in that it relies on an hybrid approach to the problem of galactic parameter estimation which is neither purely photometric nor orbit-based, as commonly employed in those sort of studies (\citealt{Mendel_2014}, \citealt{Zhu_2018b}). The algorithm we developed exploits simultaneously the entire wealth of photometric, kinematic, and mass-to-light ratio measurements available for single targets. We adopt an idealized model of disc galaxy whose structure is in the form of an axi-symmetric superposition of one or more discs and a central bulge, embedded in the host dark-matter halo. The model, described by 17 parameters, is then fitted against photometric, kinematic, and mass-to-light ratio data.  

In order to keep the galaxy parameter estimation reasonable in terms of computational burden, we made a number of necessary assumptions and simplifications, namely:
\begin{itemize}
    \item the bulge kinematics neglects the effect of the disc components on the potential. This approximation is better suited to model compact bulges, and may result in a poor description of more extended spheroidal components;
    \item the mass-to-light ratio is constant in radius for each galaxy component. Note that, once the galactic components are combined together, this results in a mass-to-light ratio that effectively depends upon the galaxy radial coordinate; 
    \item the disc component(s) is(are) assumed to be geometrically thin, its(their) kinematics is obtained by allowing the total kinetic energy to be shared between ordered and isotropic motions.   
\end{itemize}

Given the galaxy model just described, the parameter estimation is performed via a Bayesian  nested sampling analysis, that automatically evaluates the evidence of each model allowing for direct model selection. Since the analysis is intrinsically computationally demanding (e.g., a large number of evaluations of the likelihood is involved; the calculation of the evidence is regarded as simply {\it too} demanding in many applications; etc.), in order to speed-up calculations the code has been optimized for running on GPUs. The parameter estimation has been specifically tailored to perform an accurate, still reasonably fast survey of the 17-dimensional parameter space. This results in a clean determination of the posterior probability distributions of all 17 fitting parameters involved in the model, as shown in Figure~\ref{fig:corner_plot}. 

As a first case-study, we applied the algorithm to NGC~7683, 
an S0 galaxy at redshift $z \simeq 0.012$, whose kinematics has been obtained as part of the CALIFA survey \citep{Sanchez_2012}, while the \textit{ugriz}-band photometry was collected within the Sloan Digital Sky Survey \citep{SDSS}. The Bayesian analysis is able to: fully constrain the posterior probability distributions of all 17 fitting parameters; detect with high statistical significance the presence of a rotating component in the very center of NGC~7683. This may suggest that NGC~7683 possesses a rotating pseudo-bulge, possibly a bar remnant. We compared our parameter estimation with the results we obtained for the same galaxy with the publicly available code DYNAMITE \citep[][]{DYNAMITE}, a numerical tool based on orbit superposition \citep[][]{Schwarzschild1979,Bosch_2008}, finding substantial agreement. 

The present effort must be considered as a first step towards a systematic analysis of the information deriving from the kinematics, photometry and the mass-to-light ratio of disc galaxies, with the aim of deriving their underlying structure. The results obtained are indeed very promising, and make the proposed methodology an effective and available tool for estimating galactic parameters, pushing to a more comprehensive exploitation of the code. As a matter of fact, we are currently planning to utilize our code in a systematic study of a large sample of disc galaxies. This inevitably calls for a more advanced engineering in order to further reduce the computational burden. To this aim, we are working on an up-dated version of the code rooted on machine learning strategies, specifically we are implementing a strategy based on state-of-the-art neural networks for super-resolution image reconstruction \citep[see, e.g.,][]{FSRCNN}. This new effort is underway, and it will be presented in a future paper.  

\section*{Acknowledgements}
We gratefully thank the \textsc{dynamite} team, in particular Sabine Thater, for the many insightful and constructive discussions, \fr{we thank the anonymous referee for the useful comments} that helped us to improve the quality and readability of our work.
AL acknowledges funding from MIUR under the grant PRIN 2017-MB8AEZ.
\section*{Data Availability}

The data underlying this article will be shared on reasonable request to the corresponding author.



\bibliographystyle{mnras}
\bibliography{main} 




\appendix

\section{GPU parallelization}
\label{app:GPU_parallelization}

The evaluation of each model on CPU passes through two main loops. We build around each pixel a subgrid of $K$ new points where we evaluate and then convolve all the relevant quantities described in Section~\ref{sec:methodology}. Given $N_x$ and $N_y$ number of pixels in the horizontal and vertical direction, the total number of operations is $\mathcal{O}(N_x\times N_y \times K$). 

\begin{algorithm}
	\caption{Model evaluation on CPU} 
	\begin{algorithmic}[1]
		\For {$iteration=1,\ldots,N_x\times N_y$}
		    \For {$iteration=1,\ldots,K$}
				\State Evaluate $B_{tot}$,$v_{los}$,$\sigma_{los}$,$\left<M/L\right>$ and their dependencies.
				\State Average the relevant quantities weighted on the PSF.
		\EndFor
	\EndFor
	\end{algorithmic} 
\end{algorithm}

Note that, due to the grid-based nature of our model, most of the computations done on two different pixels are independent of each other and they can be easily parallelized. However, part of the code must be kept serial. For example, it is not possible to calculate the projected radius $R$ in each pixel before translating the galaxy to the coordinates center.

In this context, where a significant part of the computation may be performed in parallel, GPUs are ideal tools since they are composed of thousands of cores (threads) that can process a lot of simple operations simultaneously. GPU programming is based on the declaration of the so-called kernels, which are `special functions' that can be run in parallel on the cores of the GPU die. In this way, in each clock cycle the same kernel can be evaluated a large number of times simultaneously.
GPUs can be visualized as grids organized in blocks each containing a number of threads. The number of blocks per grid and the number of threads per block must be optimized in order to reach maximum efficiency.
In our approach, each evaluation of the model can be visualised as a series of operations on a 3 dimensional  $(N_x, N_y, K)$ tensor. Although GPU could, in principle, operate directly on multi-dimensional arrays, we modeled our tensors as 1D flattened vectors to speed up the computation thanks to a lower number of memory accesses.

The overall structure of the code once moved on GPUs can be summarized as:

\begin{algorithm}
	\caption{Model evaluation on GPU} 
	\begin{algorithmic}[1]
	    \State define a serial structure of kernels each one related to a precise task of the model (e.g. coordinates transformation, density computation, etc.).
	    \State allocate the necessary amount of memory on the GPU and upload all the CPU vectors there.
	    \State run sequentially all the kernels with $N_x \times N_y \times K$ threads in order to compute all the necessary quantities.
	    \State run the kernels dedicated to the PSF convolution ($N_x \times N_y$ threads summing over $K$).
	    \State download from the GPU (if necessary) the output of the model. 
	\end{algorithmic} 
\end{algorithm}
We fixed the number of threads per block to 64 while the number of block per grid depends on the total number of threads required by a specific kernel.


The memory allocation and the loading operation are performed only once at the beginning of the parameter estimation. This is crucial in order to reach a relevant speed up since loading operations are usually severe bottlenecks for GPUs programming. With the GPU porting we finally reduced the computational cost of each single model evaluation by $210$ times. 

\section{Jaffe profile}
\label{app:Jaffe_profile}
The model with 3-D density  
\begin{equation}
     \rho_{\rm b}(r) = \frac{M_{\rm b}}{4\pi}\frac{R_{\rm b}}{r^2(r+R_{\rm b})^2}  
     \label{eq:Jaffe_density}
\end{equation}
refers to the Jaffe profile which, similarly to the Herquinst model, is commonly used to describe bulges and elliptical galaxies.  

The circular velocity determined by the Jaffe potential is   
 \begin{equation}
     v_{\mathrm{circ,b}}(r) = \sqrt{\frac{GM_{\rm b}}{r+R_{\rm b}}}; 
     \label{eq:Jaffe_circular_velocity}
 \end{equation}
while the radial velocity dispersion reads
 \begin{equation}
     \begin{split}
     \overline{v_{r}^2}(r) =& \frac{GM_{\rm b}r^2(r+R_{\rm b})^2}{a^4}\Biggl\{\frac{3}{r}+\frac{6r+7R_{\rm b}}{2(R_{\rm b}+r)^2}\\&-\frac{R_{\rm b}}{2r^2}-\frac{6}{R_{\rm b}}\log{\left(1+\frac{R_{\rm b}}{a}\right)}\Biggr\} .\end{split}
     \label{eq:Jaffe_velocity_dispersion}
 \end{equation}



The surface brightness and the line of sight velocity dispersion for the isotropic Jaffe model are analytical and correspond to
 \begin{equation}
     \Sigma_{\rm b}(R) = \frac{M_{\rm b}}{R_{\rm b}^2}\left[\frac{1}{4s}+\frac{1-(2-s^2)X(s)}{2\pi(1-s^2)}\right],
     \label{eq:Jaffe_projected_density}
 \end{equation}
and
 \begin{equation}
 \begin{split}
  \sigma_{\rm b}^2(R) = \frac{GM_{\rm b}^2}{4 \upi R_{\rm b}^3 \Sigma_{\rm b}(R)}\frac{1}{\left(1-s^2\right)}\Biggl[\frac{\upi}{2s}+11-\frac{13}{2}\upi s -12s^2 + 6\upi s^3 \\ -X(s)\left(6-19s^2+12s^4\right)\Biggr]. \end{split}
   \label{eq:Jaffe_projected_dispersion}
 \end{equation}
 respectively.

\section{$M/L$ ratio from colors-$M/L$ relation}
\label{app:ML_from_color}
We report here the same analysis performed in sec.~\ref{sec:our_analysis} while considering $M/L$ data computed from eq.~\eqref{eq:ML_ratio}. More specifically, in the case of the NGC 7683 galaxy we choose $\alpha=\gamma$ corresponding to the SDSS i-band  and $\beta$ as the SDSS g-band. We underline that the $M/L$-color relation is only a proxy for the real mass-to-light ratio and relies on several assumptions such as the star formation history and the initial mass function, assumptions that we take into account when computing the associated errors. More precisely, the uncertainty on the output of Eq.~\eqref{eq:ML-color} is computed by propagating the error on the magnitudes $m_{\beta}$, $m_{\gamma}$ and adding in quadrature a tabulated constant\footnote{See the appendices of \citealt{Garcia_2018}.} $\sigma_\alpha$ which account for the scatter in the color--$M/L$ relation. In our test case we fixed the tabulated parameters $a_\alpha=-0.70$, $b_\alpha=0.89$, and $\sigma_\alpha=0.07$.
\begin{table}
	\centering
	\caption{Same as tab.~\ref{tab:best_fit_parameters} using $M/L$ ratio data from $M/L$-color relation. }
	\label{tab:best_fit_parameters_app}
	\begin{tabular}{lcr} 
		\hline
		Parameter & Prior range & Best fit value \\
		\hline \rowsep
		$x_0 [\mathrm{kpc}] $ &  $[-2,1]$ & $-1.031^{+0.002}_{-0.002}$ \\ \rowsep
		$y_0 [\mathrm{kpc}]$ &  $[-2,1]$ & $-1.055^{+0.002}_{-0.002}$\\  \rowsep
		$\rm P.A. [\mathrm{deg}]$ &  $[-180,180]$ & $44.12^{0.06}_{0.06}$ \\ \rowsep
		$\sin{i}$ &  $[0.17,0.97]$ & $0.827^{0.001}_{0.001}$\\ \rowsep
		$\log_{10}(M_{\rm b}/\rm M_{\sun}) $ &  $[8.5;11.5]$ & $10.64^{0.01}_{0.01}$\\ \rowsep
		$\log_{10}(R_{\rm b}/\mathrm{kpc}) $ &  $[-2,1]$ & $-0.71^{0.01}_{0.01}$ \\ \rowsep
		$\log_{10}(M_{\rm d,1}/\rm M_{\sun})$ &  $[8.5;11.5]$ & $10.40^{0.01}_{0.01}$ \\  \rowsep
		$\log_{10}(R_{\rm d,1}/\mathrm{kpc}) $ &  $[-2,1]$ & $-0.029^{0.003}_{0.003}$ \\ \rowsep
		$\log_{10}(M_{\rm d,2}/\rm M_{\sun})$ &  $[8.5;11.5]$ & $10.860^{0.002}_{0.002}$ \\ \rowsep
		$\log_{10}(R_{\rm d,2}/\mathrm{kpc}) $ &  $[-2,1]$ & $0.597^{0.002}_{0.002}$ \\ \rowsep
		$\log_{10}(M_{\rm h}/\rm M_{\sun})$ &   $[11.5,14.5]$ & - \\ \rowsep
		$\log_{10}(R_{\rm h}/\mathrm{kpc})$ &  $[1;3]$ & - \\ \rowsep
		$\log_{10}{M_{\rm h,5}/\rm M_{\sun}}$ & - & $11.870^{0.006}_{0.008}$ \\ \rowsep
		$(M/L)_{\rm b} [\rm{L_{\sun}/M_{\sun}}]$ &  $[1;100]$ & $9.90^{0.2}_{0.2}$ \\ \rowsep
		$(M/L)_{\rm d,1} [\mathrm{L_{\sun}/M_{\sun}}]$ &  $[0.1;100]$ & $1.96^{0.02}_{0.02}$ \\ \rowsep
		$(M/L)_{\rm d,2} [\mathrm{L_{\sun}/M_{\sun}}]$ &  $[0.1;100]$ & $2.58^{0.01}_{0.01}$ \\ \rowsep
		$k_1$ &  $[0;1]$ & $0.302^{0.002}_{0.002}$ \\ \rowsep
		$k_2$ &  $[0;1]$ & $0.695^{0.001}_{0.001}$ \\
		\hline
	\end{tabular}
\end{table}

Comparing the best fit parameters obtained in the two different cases (tab.~\ref{tab:best_fit_parameters} and tab.~\ref{tab:best_fit_parameters_app}) we can see that they are all recovered within a maximum relative discrepancy of a few percents, see for example that the total visible masses respectively of  $1.41^{+0.01}_{-0.01}\times10^{11}M_{\sun}$ in this case and $1.46^{+0.02}_{-0.02}\times10^{11}M_{\sun}$ in the SPS $M/L$ ratio case which are consistent within $2\sigma$ uncertainty. Based on this comparison we can conclude that the two method are compatible, even though we suggest, when possible, to prefer SPS modelling due to is better accuracy in predicting $M/L$ ratio data.  

\bsp	
\label{lastpage}
\end{document}